\documentclass{article}
\usepackage{amssymb}
\usepackage{amsmath}
\usepackage{amsfonts}
\usepackage{hyperref}
\usepackage{multimedia}
\usepackage{graphicx}
\usepackage{setspace}
\usepackage{subfigure}
\usepackage{rotating}
\usepackage{booktabs}

\input{tcilatex}
\begin{document}

\title{Modern Monetary Circuit Theory, Stability of Interconnected Banking
Network, and Balance Sheet Optimization for Individual Banks}
\author{Alexander Lipton\footnote{The views and opinions expressed in this paper are those of the author and
do not necessarily reflect the views and opinions of Bank of America.}  \\
Bank of America and University of Oxford}
\maketitle

\begin{abstract}
A modern version of Monetary Circuit Theory with a particular emphasis on
stochastic underpinning mechanisms is developed. It is explained how money
is created by the banking system as a whole and by individual banks. The
role of central banks as system stabilizers and liquidity providers is
elucidated. It is shown how in the process of money creation banks become
naturally interconnected. A novel Extended Structural Default Model
describing the stability of the Interconnected Banking Network is proposed.
The purpose of banks' capital and liquidity is explained. Multi-period
constrained optimization problem for banks's balance sheet is formulated and
solved in a simple case. Both theoretical and practical aspects are covered.
\end{abstract}

\tableofcontents

.\pagebreak
\begin{verbatim}
Steven Obanno: Do you believe in God, Mr. Le Chiffre?
Le Chiffre: No. I believe in a reasonable rate of return.
Casino Royale
 
Coffee Cart Man: Hey buddy. You forgot your change.
Joe Moore: [Takes the change] Makes the world go round.
Bobby Blane: What's that?
Joe Moore: Gold.
Bobby Blane: Some people say love.
Joe Moore: Well, they're right, too. It is love. Love of gold.
Heist
\end{verbatim}

\section{Introduction\label{Introduction}}

Since times immemorial, the meaning of money has preoccupied industrialists,
traders, statesmen, economists, mathematicians, philosophers, artists, and
laymen alike.

The great British economist John Maynard Keynes puts it succinctly as
follows:

\begin{quote}
For the importance of money essentially flows from it being a link between
the present and the future.
\end{quote}

These words are echoed by Mickey Bergman, the character played by Danny
DeVito in the movie Heist, who says:

\begin{quote}
Everybody needs money. That's why they call it money.
\end{quote}

Money has been subject of innumerable expositions, see, e.g., Law (1705),
Jevons (1875), Knapp (1905), Schlesinger (1914), von Mises (1924), Friedman
(1969), Schumpeter (1970), Friedman and Schwartz (1982), Kocherlakota
(1998), Realfonzo (1998), Mehrling (2000), Davidson (2002), Ingham (2004),
Graeber (2011), McLeay \textit{et al}. (2014), among many others. Recently,
these discussions have been invigorated by the introduction of Bitcoin
(Nakamoto 2009). An astute reader will recognize, however, that apart from
intriguing technical innovations, Bitcoin does not differ that much from the
fabled tally sticks, which were used in the Middle Ages, see, e.g., Baxter
(1989). It is universally accepted that money has several important
functions, such as a store of value, a means of payment, and a unit of
account.\footnote{%
We emphasize that a particularly important function of money as a means of
payment of \emph{taxes}.}

However, it is extraordinary difficult to understand the role played by
money and to follow its flow in the economy. One needs to account properly
for non-financial and financial stocks (various cumulative amounts), and
flows (changes in these amounts). Here is how Michal Kalecki, the great
Polish economist, summarizes the issue with his usual flair and penchant for
hyperbole:

\begin{quote}
Economics\ is\ the\ science\ of\ confusing\ stocks\ with\ flows.
\end{quote}

In our opinion, the functioning of the economy and the role of money is best
described by the Monetary Circuit Theory (MCT), which provides the framework
for specifying how money lubricates and facilitates production and
consumption cycles in society. Although the theory itself is quite
established, it fails to include some salient features of the real economy,
which came to the fore during the latest financial crisis. The aim of the
current paper is to develop a modern continuous time version of this
venerable theory, which is capable of dealing with the equality between
production and consumption plus investment, the stochastic nature of
consumption, which drives other economic variables, defaults of the
borrowers, the finite capacity of the banking system for lending, etc. This
paper provides a novel description of the behaviour and stability of the
interlinked banking system, as well as of the role played by individual
banks in facilitating the functioning of the real economy. The latter aspect
is particularly important because currently there is a certain lack of
appreciation on the part of the conventional economic paradigm of the
special role of banks. For example, banks are excluded from widely used
dynamic stochastic general equilibrium models, which are presently
influential in contemporary macroeconomics (Sbordone \textit{et al}. 2010).

Some of the key insights on the operation of the economy can be found in
Smith (1776), Marx (1867), Schumpeter (1912), Keynes (1936), Kalecki (1939),
Sraffa (1960), Minsky (1975, 1986), Stiglitz (1997), Tobin \& Golub (1998),
Piketty (2014), Dalio (2015), etc. The reader should be cognizant of the
fact that opinions of the cited authors very often contradict each other, so
that the "correct" viewpoint on the actual functioning of the economy is not
readily discernible.

Monetary Circuit Theory, which can be viewed as a specialized branch of the
general economic theory, has a long history. Some of the key historical
references are Petty\ (1662), Cantillon (1755), Quesnay (1759), Jevons
(1875). More recently, this theory has been systematically developed by Keen
(1995, 2013, 2014) and others. The theory is known under several names such
as Stock-Flow Consistent (SFC) Model, Social Accounting Matrix (SAM) Model,
etc. Post-Keynsian SFC macroeconomic growth models are discussed in numerous
references. Here is a representative selection: Backus et al. (1980), Tobin
(1982), Moore (1986, 2006), De Carvalho (1992), Godley (1999), Bellofiore 
\textit{et al}. (2000), Parguez and Secareccia (2000), Lavoie (2001, 2004),
Lavoie and Godley (2001-2002), Gnos (2003), Graziani (2003), Secareccia
(2003), Dos Santos and Zezza (2004, 2006), Zezza \& Dos Santos (2004),
Godley and Lavoie (2007), Van Treek (2007), Le Heron and Mouakil (2008), Le
Heron (2009), Dallery and van Treeck (2011). A useful survey of some recent
results is given by Caverzasi and Godin (2013).

It is a simple statement of fact that reasonable people can disagree about
the way money is created. Currently, there are three prevailing theories
describing the process of money creation. Credit creation theory of banking
has been dominant in the 19th and early 20th centuries. It is discussed in a
number of books and papers, such as Macleod (1855-6), Mitchell-Innes (1914),
Hahn (1920), Wicksell (1922), and Werner (2005). More recently Werner (2014)
has empirically illustrated how a bank can individually create money "out of
nothing"\footnote{%
However, his experiment was not complete because he received a loan from the
same bank he has deposited the money to. As discussed later, this is a very
limited example of monetary creation.}. In our opinion, this theory
correctly reflects mechanics of linking credit and money creation;
unfortunately, it has gradually lost its ground and was overtaken by the
fractional reserve theory of banking, see for example, Marshall (1888),
Keynes (1930), Samuelson \& Nordhaus (1995), and numerous other sources.
Finally, the financial intermediation theory of banking is the current
champion, three representative descriptions of this theory are given by
Keynes (1936), Tobin (1969), and Bernanke \& Blinder (1989), among many
others. In our opinion, this theory puts insufficient emphasis on the unique
and special role of the banking sector in the process of money creation.

In the present paper we analyze the process of money creation and its
intrinsic connection\textbf{\ }to credit in the modern economy. In
particular, we address the following important questions: (a) Why do we need
banks and what is their role in society? (b) Can a financial system operate
without banks? (c) How do banks become interconnected as a part of their
regular lending activities? (d) What makes banks different from
non-financial institutions? In addition, we consider a number of issues
pertinent to individual banks, such as (e) How much capital do banks need?
(f) How liquidity and capital are related? (g) How to optimize a bank
balance sheet? (h) How would an ideal bank look like? (i) What are the
similarities and differences between insurance companies and banks viewed as
dividend-producing machines? In order to answer these crucial questions we
develop a new Modern Monetary Circuit (MMC) theory, which treats the banking
system on three levels: (a) the system as a whole; (b) an interconnected set
of individual banks; (c) individual banks. We try to be as parsimonious as
possible without sacrificing an accurate description of the modern economy
with a particular emphasis on credit channels of money creation in the
supply-demand context and their stochastic nature.

The paper is organized as follows. Initially, in Sections \ref{LVGE} and \ref%
{KE} we develop the building blocks, which are further aggregated in Section %
\ref{SimpleEconomy} into the consistent continuous time MMC\textbf{\ }%
theory. In Section \ref{LVGE}, we introduce stochasticity into conventional
Lotka-Volterra-Goodwin equations and incorporate natural restrictions on the
relevant economic variables. Further, in Section \ref{KE} we analyze the
conventional Keen equations and modify them by incorporating stochastic
effects and natural boundaries. Building upon the results of Sections \ref%
{LVGE} and \ref{KE}, we develop in Section \ref{SimpleEconomy} a consistent
MMC\textbf{\ }theory and illustrate it for a simple economic triangle that
includes\textbf{\ }consumers (workers and rentiers), producers and banks.
Section \ref{MoneyCreation} details the underlying process of money creation
and annihilation by the banking system and discusses the role of the central
bank as the liquidity provider for individual banks. In Section \ref%
{BankingSystem} we develop the framework to study the banking system, which
becomes interlinked in the process of money creation and propose an extended
structural default model for the interconnected banking network. This model
is further explained in Appendix A for the simple case of two interlinked
banks with mutual obligations. In Section \ref{BSOptimization} the behaviour
of individual banks operating as a part of the whole banking system is
analyzed with an emphasis on the role of banks' capital and liquidity. The
balance sheet optimization problem for an individual bank is formulated and
solved in a simplified\textbf{\ }case.

\section{Stochastic Modified Lotka-Volterra-Goodwin Equations\label{LVGE}}

\subsection{Background}

The Lotka-Volterra system of first-order non-linear differential equations
qualitatively describes the predator-prey dynamics observed in biology {%
(Lotka 1925, Volterra 1931)}. Goodwin was the first to apply these equations
to the theory of economic growth and business cycles {(Goodwin 1967).} His
equations, which establish\textbf{\ }the relationship between the worker's
share of national income and employment rate became deservedly popular
because of their simple and parsimonious nature and ability to provide a
qualitative description of the business cycle. However, they do have several
serious drawbacks, including their non-stochasticity, prescriptive nature of
firms' investment decisions, and frequent violations of natural restrictions
on the corresponding economic variables. Although, multiple extensions of
the Goodwin theory have been developed over time (see, e.g., Solow 1990,
Franke \textit{et al}. 2006, Barbosa-Filho and Taylor 2006, Veneziani and
Mohun 2006, Desai \textit{et al}. 2006, Harvie \textit{et al.} 2007, Kodera
and Vosvrda 2007, Taylor 2012, Huu and Costa-Lima 2014, among others), none
of them is able of holistically account for all the deficiencies outlined
above. In this section we propose a novel mathematically consistent version
of the Goodwin equations, which we subsequently use as a building block for
the MMC theory described in Section \ref{SimpleEconomy}.

\subsection{Framework}

Assume, for simplicity, that in the stylized economy a single good is
produced.\textbf{\ }Then the productivity of labor $\theta _{w}$ is measured
in production units per worker per unit of time, the available workforce $%
N_{w}$ is measured in the number of workers, while the employment rate $%
\lambda _{w}$ is measured in fractions of one. Thus, the total number of
units produced by firms per unit of time, $\Upsilon _{f}$, is given by%
\begin{equation}
\Upsilon _{f}=\theta _{w}\lambda _{w}N_{w},  \label{Eq1}
\end{equation}%
where both productivity and labor pool grow deterministically as%
\begin{equation}
\frac{d\theta _{w}}{\theta _{w}}=\alpha dt,  \label{Eq2}
\end{equation}%
\begin{equation}
\frac{dN_{w}}{N_{w}}=\beta dt.  \label{Eq3}
\end{equation}%
If so desired, these relations can be made much more complicated, for
example, we can add stochasticity, more realistic population dynamics, etc.
Production expressed in monetary terms is given by%
\begin{equation}
Y_{f}=\theta _{w}\lambda _{w}N_{w}P,  \label{Eq4}
\end{equation}%
where $P$ is the price of one unit of goods. Similarly to Eqs (\ref{Eq2}), (%
\ref{Eq3}) we assume that the price is deterministic, such that 
\begin{equation}
\frac{dP}{P}=\gamma dt.  \label{Eq5}
\end{equation}%
Workers' and firms' share of production are denoted by $s_{w},s_{f}=1-s_{w}$%
, respectively. The unemployment rate $\lambda _{u}$ is defined in the usual
way, $\lambda _{u}=1-\lambda _{w}$. Goodwin's idea was to describe joint
dynamics of the pair $\left( s_{w},\lambda _{w}\right) $.

\subsection{Existing Theory}

The non-stochastic {Lotka-Volterra-Goodwin equations (LVGEs, see Lotka 1925,
Volterra 1931, Goodwin 1967),} describe the relation between the workers'
portion of the output and the relative employment rate.

The log-change of $s_{w}$ is govern by the Phillips law and can be written
in the form%
\begin{equation}
\frac{ds_{w}}{s_{w}}=\left( -a+b\lambda _{w}\right) dt\equiv \phi \left(
\lambda \right) dt,  \label{Eq8}
\end{equation}%
where $\phi \left( \lambda \right) $ it the so-called Phillips curve
(Phillips 1958, Flaschel 2010, Blanchflower and Oswald 1994).

The log-change of $\lambda _{w}$ is calculated in three easy steps. First,
the so-called Cassel-Harrod-Domar (see Cassel 1924, Harrod 1939, Domar 1946)
law is used to show that%
\begin{equation}
Y_{f}=\nu _{f}K_{f},  \label{Eq9}
\end{equation}%
where $K_{f}$\ is the monetary value of the firm's non-financial assets and $%
\nu _{f}$ is the constant production rate, which is the inverse of the
capital-to-output ratio $\varpi _{f}$, $\nu _{f}=1/\varpi _{f}$.\footnote{%
In essence, we apply the celebrated Hooke's law (\textit{ut tensio, sic vis)}
in the economic context.} It is clear that $\nu _{f}$, which can be thought
of as a rate, is measured in units of inverse time, $\left[ 1/T\right] $,
while $\varpi _{f}$\ is measured in units of time, $\left[ T\right] $.
Second, Say's law (Say 1803), which states that all the firms' profits,
given by 
\begin{equation}
\Pi _{f}=s_{f}Y_{f}=s_{f}\nu _{f}K_{f},  \label{Eq10}
\end{equation}%
are re-invested into business, so that the dynamics of $K_{f}$ is govern by
the following deterministic equation%
\begin{equation}
\frac{dK_{f}}{K_{f}}=\frac{dY_{f}}{Y_{f}}=\left( s_{f}\nu _{f}-\xi
_{A}\right) dt,  \label{Eq11}
\end{equation}%
with $\xi _{A}$ being the amortization rate. Finally, the relative change in
employment rate, $\lambda _{w}$ is derived by combining Eqs. (\ref{Eq2}) - (%
\ref{Eq5}) and (\ref{Eq11}):%
\begin{equation}
\frac{d\lambda _{w}}{\lambda _{w}}=\frac{dY_{f}}{Y_{f}}-\frac{d\theta _{w}}{%
\theta _{w}}-\frac{dN_{w}}{N_{w}}-\frac{dP}{P}=\left( s_{f}\nu _{f}-\alpha
-\beta -\gamma -\xi _{A}\right) dt.  \label{Eq12}
\end{equation}%
Symbolically,%
\begin{equation}
\frac{d\lambda _{w}}{\lambda _{w}}=\left( c-ds_{w}\right) dt.  \label{Eq13}
\end{equation}%
Thus, the coupled system of equations for $\left( s_{w},\lambda _{w}\right) $
has the form%
\begin{eqnarray}
\frac{ds_{w}}{s_{w}} &=&-\left( a-b\lambda _{w}\right) dt,  \label{Eq17} \\
\frac{d\lambda _{w}}{\lambda _{w}} &=&\left( c-ds_{w}\right) dt.  \notag
\end{eqnarray}

\begin{sidewaysfigure}[t]
\center%
\includegraphics[width=1.0\textwidth, angle=0]{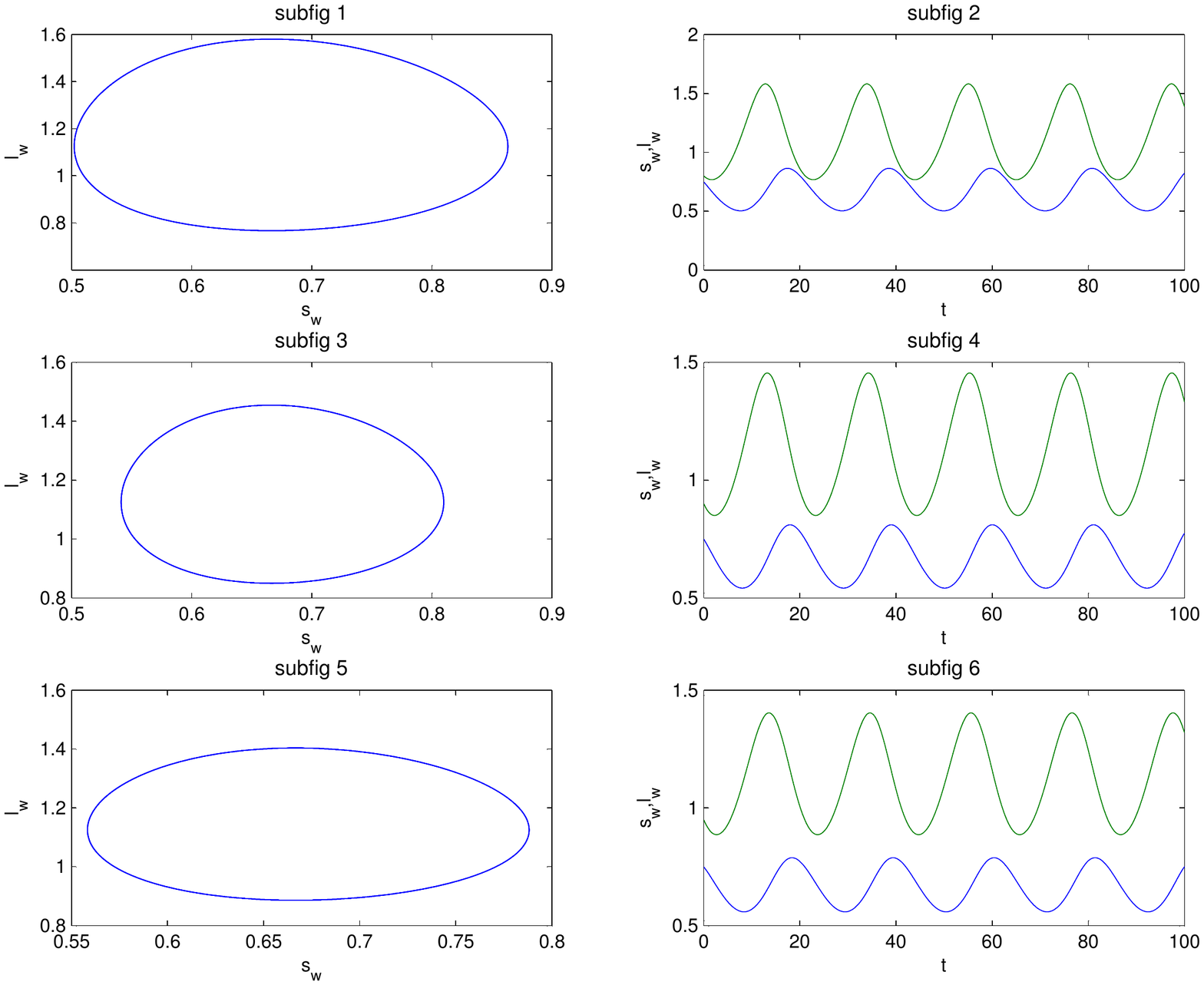}
\caption{{}Typical solutions of LVGEs without regularization. Natural
constraints are violated. Representative parameters are $a=0.225, b=0.20, c=0.4, d=0.6$. 
Initial conditions for subfigures 1-2, 3-4, and 5-6 are $(s_w=0.75,\lambda_w=0.8)$,
$(s_w=0.75,\lambda_w=0.9)$, and $(s_w=0.75,\lambda_w=0.95)$, respectively.}
\label{Fig1}
\end{sidewaysfigure}

Eqs (\ref{Eq17}) schematically describe the class struggle; they are
formally identical to the famous predator-pray Lotka-Volterra equations in
biology, with intensive variables $s_{w},\lambda _{w}$ playing the role of
predator and pray, respectively. Two essential drawbacks of the LGVE are
that they neglect the stochastic nature of economic processes and do not
preserve natural constraints $\left( s_{w},\lambda _{w}\right) \in \left(
0,1\right) \times \left( 0,1\right) $. Besides, they are too restrictive in
describing the discretionary nature of firms' investment decisions. {The
conservation law }$\Psi ${\ corresponding Eqs. (\ref{Eq17}) has the
following form}%
\begin{equation}
{\Psi \left( s_{w},\lambda _{w}\right) =-\ln \left( s_{w}^{c}\lambda
_{w}^{a}\right) +ds_{w}+b\lambda _{w},}  \label{Eq14}
\end{equation}%
and has a {fixed point at }%
\begin{equation}
\left( \frac{c}{d},\frac{a}{b}\right) ,  \label{Eq15}
\end{equation}%
{where $\Psi $ achieves its minimum. }\QTR{frametitle}{Solutions of the
LVGEs without regularization are shown in Figure \ref{Fig1}. Both phase
diagrams in the }${\left( s_{w},\lambda _{w}\right) }$\QTR{frametitle}{%
-space and time evolution graphs show that for the chosen set of parameters }%
$\lambda _{w}>1$\QTR{frametitle}{\ for some parts of the cycle.}

\begin{equation*}
\text{Figure \ref{Fig1} near here.}
\end{equation*}

\subsection{Modified Theory}

\begin{sidewaysfigure}[t]
\center%
\includegraphics[width=1.0\textwidth, angle=0]{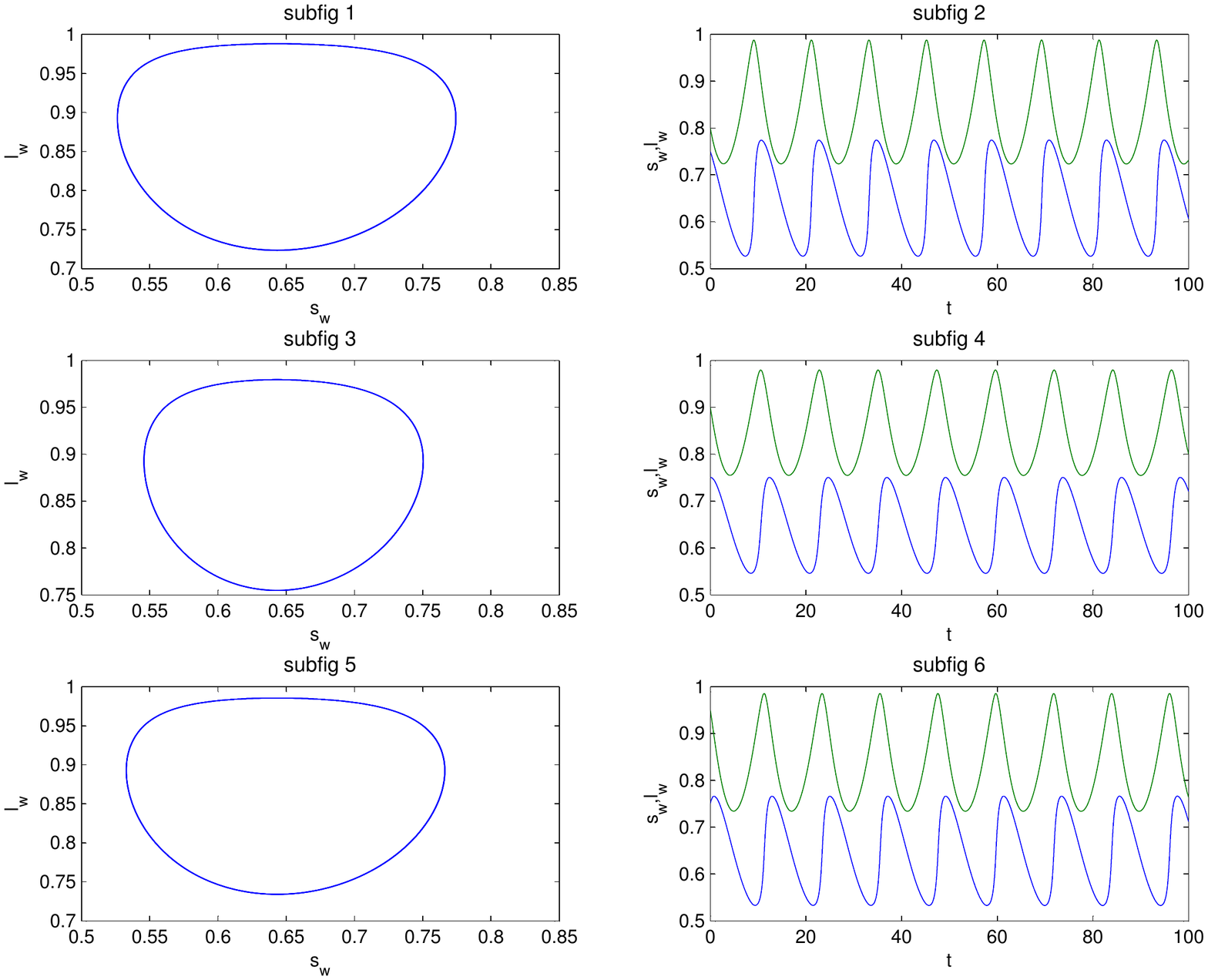}
\caption{{}Typical solutions of LVGEs with regularization. Natural
constraints are satisfied. The same parameters and initial conditions as in Figure \ref{Fig1} are used;
in addition, $\omega=0.005$.}
\label{Fig2}
\end{sidewaysfigure}

\begin{sidewaysfigure}[t]
\center%
\includegraphics[width=1.0\textwidth, angle=0]{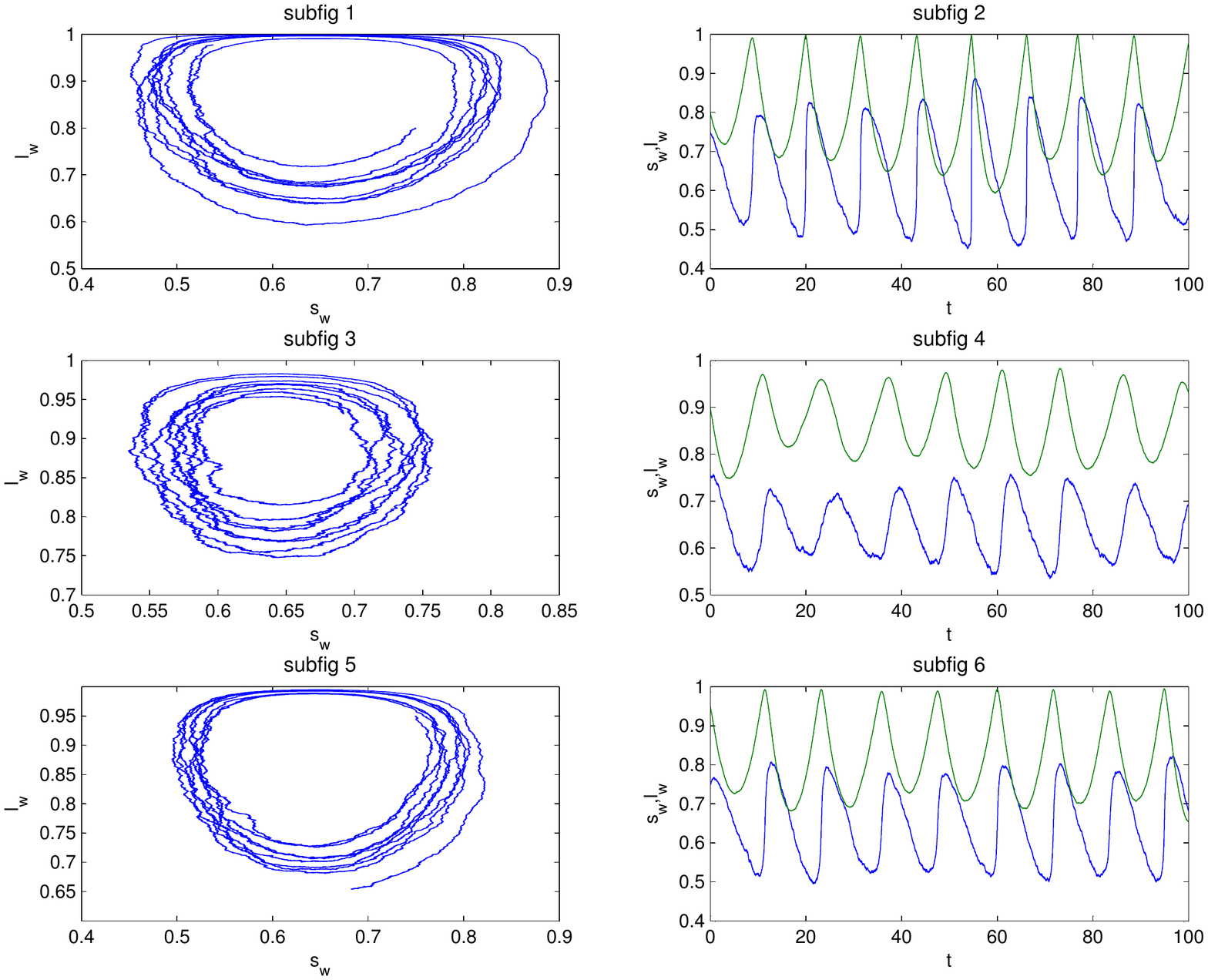}
\caption{{}Typical solutions of LVGEs with regularization and stochasticity. Natural
constraints are satisfied; orbits are non-periodic. 
The same parameters and initial conditions as in Figure \ref{Fig2} are used;
in addition, $\sigma_s=0.015, \sigma_l=0.005$.}
\label{Fig3}
\end{sidewaysfigure}

In order to satisfy natural boundaries in the stochastic framework, we
propose {a new version of the \QTR{frametitle}{LVGEs} of the form%
\begin{eqnarray}
ds_{w} &=&-\left( a-b\lambda _{w}-\frac{\omega }{\lambda _{u}}\right)
s_{w}dt+\sigma _{s}\sqrt{s_{w}s_{f}}dW_{s}\left( t\right) ,  \label{Eq16} \\
d\lambda _{w} &=&\left( c-ds_{w}-\frac{\omega }{s_{f}}\right) \lambda
_{w}dt+\sigma _{\lambda }\sqrt{\lambda _{w}\lambda _{u}}dW_{\lambda }\left(
t\right) ,  \notag
\end{eqnarray}%
where }$\omega >0$ is a regularization parameter{\small , }and $\sigma _{s}%
\sqrt{s_{w}s_{f}}$, $\sigma _{\lambda }\sqrt{\lambda _{w}\lambda _{u}}$\ are
Jacobi normal volatilities. This choice of volatilities ensures that ${%
\left( s_{w},\lambda _{w}\right) }$\ stays within the unit square. {%
Deterministic conservation law }$\Psi ${\ }for Eqs{\textbf{\ (\ref{Eq16}) }%
is similar to Eq. (\ref{Eq14}):}%
\begin{equation}
{\Psi \left( s_{w},\lambda _{w}\right) =-\ln \left( s_{w}^{c-\omega
}s_{f}^{\omega }\lambda _{w}^{a-\omega }\lambda _{u}^{\omega }\right)
+ds_{w}+b\lambda _{w}.}  \label{Eq18}
\end{equation}%
However, it is easy to see that the corresponding contour lines stay within
the unit square, $\left( s_{w},\lambda _{w}\right) \in \left( 0,1\right)
\times \left( 0,1\right) $. The {fixed point, where $\Psi $ achieves its
minimum, is given by}%
\begin{equation}
\left( \frac{1}{2d}\left( c+d-\sqrt{\left( c-d\right) ^{2}+4d\omega }\right)
,\frac{1}{2b}\left( a+b-\sqrt{\left( a-b\right) ^{2}+4b\omega }\right)
\right) .  \label{Eq19}
\end{equation}%
\QTR{frametitle}{Effects of regularization and effects of stochasticity
combined with regularization are shown in Figures \ref{Fig2}\thinspace\ and %
\ref{Fig3}, respectively. }It is clear that, by construction, Eqs.(\ref{Eq16}%
) reflect naturally occurring stochasticity of the corresponding economic
processes, while preserving natural bounds for $s_{w}$\ and $\lambda _{w}$.

\begin{equation*}
\text{Figure \ref{Fig2} near here.}
\end{equation*}

\begin{equation*}
\text{Figure \ref{Fig3} near here.}
\end{equation*}

The idea of regularizing the Goodwin equations was originally proposed by
Desai \textit{et al}. (2006). Our choice of the regularization function is
different from theirs but is particularly convenient for further development
and advantageous because of its parsimony. At the same time, while
stochastic LVEs are rather popular in the biological context, see, e.g., Cai
and Lin (2004), stochastic aspects of the LVGEs remain relatively
unexplored, see, however, Kodera and Vosvrda (2007), and, more recently, Huu
and Costa-Lima (2014).

\section{Stochastic Modified Keen Equations\label{KE}}

\subsection{Background}

LVGEs and their simple modifications generate phase portraits, which are
either closed or almost closed, as presented in Figures \QTR{frametitle}{\ref%
{Fig1}, \ref{Fig2}, \ref{Fig3}. Accordingly, they }can not describe unstable
economic behaviour. However, historical experience suggests that capitalist
economies are periodically prone to crises, as is elucidated by the famous
Financial Instability Hypothesis of Minsky (Minsky 1977, 1986). His theory
bridges macroeconomics and finance and, if not fully develops, then, at
least clarifies the role of banks and, more generally, debt in modern
society. Although Minsky's own attempts to formulate the theory in a\textbf{%
\ }quantitative rather than qualitative form were unsuccessfull, it was
partially accomplished by Steven Keen (Keen 1995)\textbf{. }Keen extended
the Goodwin model by abandoning its key assumption that investment is equal
to profit. Instead, he assumed that, when profit rate is high, firms invest
more than their retained earnings by borrowing from banks and vice versa.

Below we briefly discuss the Keen equations and show how to modify them in
order to remove some of their intrinsic deficiencies.

\subsection{Keen Equations}

{The Keen equations (KEs) (Keen 1995), describe the relation between the
workers' portion of the output }$s_{w}$, {the employment rate }$\lambda _{w}$%
{, }and the firms' debt $D_{f}$ relative to their non-financial assets $%
K_{f} $, $\Gamma _{f}=D_{f}/K_{f}$.\footnote{%
We deviate from the original Keen's definitions somewhat for the sake of
uniformity.} All these quantities are non-dimensional. KEs can be used to
provide quantitative description of Minsky's Financial Instability
Hypothesis (Minsky 1977).\ 

Keen expanded the Goodwin framework by abandoning one of its key
simplifications, namely, the assumption that investment equals profit.
Instead, he allowed investments to be financed by banks. This important
extension enables the description of ever increasing firms' leverage until
the point when their debt servicing becomes infeasible and an economic
crisis occurs. Subsequently, Keen (2013, 2014) augmented his original
equations in order to account for flows of funds among firms, banks, and
households. However, KEs and their extensions do not take into account the
possibility of default by borrowers, and do not reflect the fact that the
banking system's lending ability is restricted by its capital capacity. Even
more importantly, extended KEs do not explicitly guarantee that production
equals consumption plus investment. In addition, as with LVGEs, KEs do not
reflect stochasticity of the underlying economic behaviour and violate
natural boundaries. Accordingly, a\textbf{\ }detailed description of the
crisis in the Keen framework is not possible.

\begin{sidewaysfigure}[t]
\center%
\includegraphics[width=1.0\textwidth, angle=0]{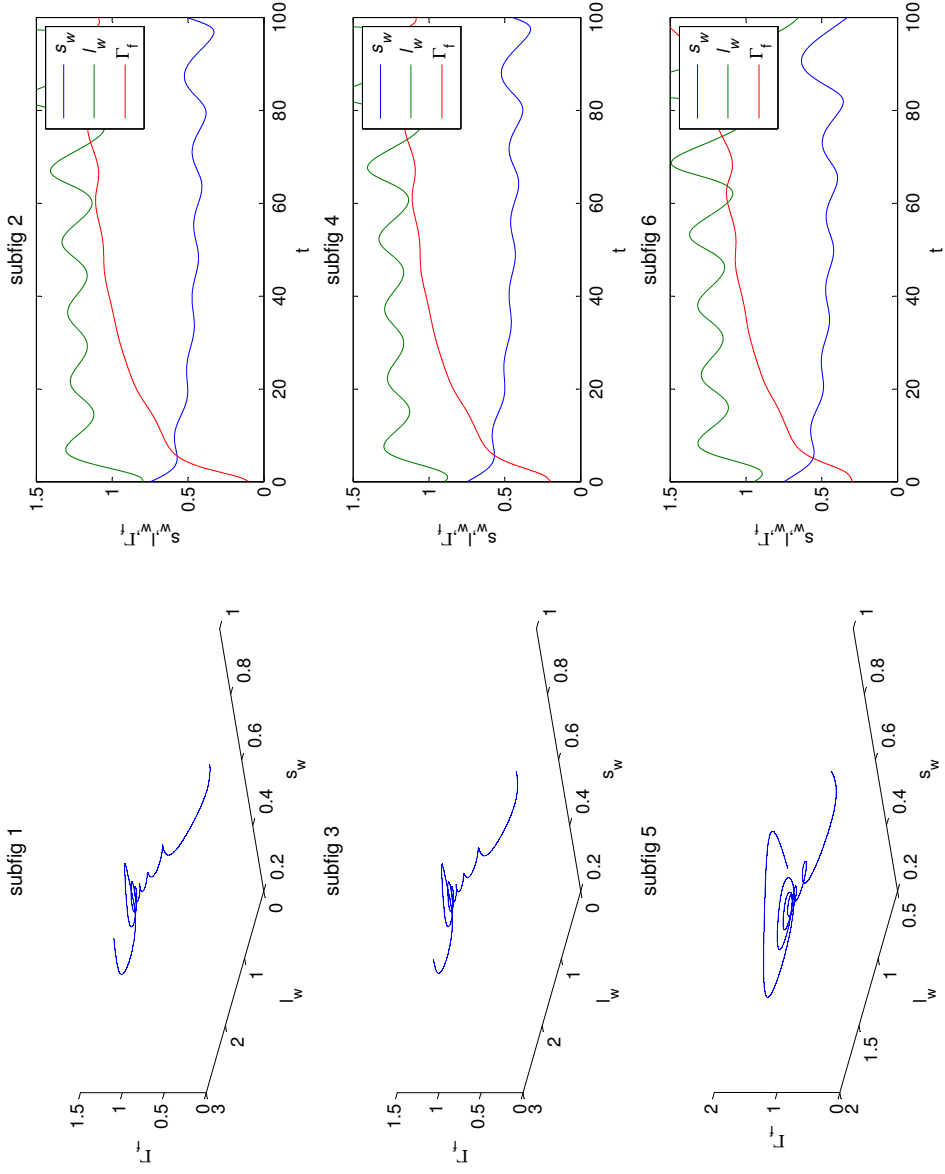}
\caption{{}Typical solutions of KEs without
regularization. Natural constraints are violated.
Representative parameters are $a=0.225, b=0.20, c=0.075, d=0.03, r_L=0.03, \nu_f=0.1,p=-0.0065, q=20.0, r=-5.0$. 
Initial conditions for subfigures 1-2, 3-4, and 5-6 are $(s_w=0.75,\lambda_w=0.8,\Gamma_f=0.1)$,
 $(s_w=0.75,\lambda_w=0.9,\Gamma_f=0.2)$, and $(s_w=0.75,\lambda_w=0.95,\Gamma_f=0.3)$, respectively.}
\label{Fig4}
\end{sidewaysfigure}

Symbolically, KEs can be written as%
\begin{eqnarray}
ds_{w} &=&-\left( a-b\lambda _{w}\right) s_{w}dt,  \label{Eq20} \\
d\lambda _{w} &=&\left( \nu _{f}f\left( s_{f}-\frac{r_{L}\Gamma _{f}}{\nu
_{f}}\right) -c\right) \lambda _{w}dt,  \notag \\
d\Gamma _{f} &=&\left( \left( r_{L}-\nu _{f}f\left( s_{f}-\frac{r_{L}\Gamma
_{f}}{\nu _{f}}\right) +d\right) \Gamma _{f}+\nu _{f}\left( f\left( s_{f}-%
\frac{r_{L}\Gamma _{f}}{\nu _{f}}\right) -s_{f}\right) \right) dt.  \notag
\end{eqnarray}%
where $a,b,c,d$ are suitable parameters, and $f\left( .\right) $ is an
increasing function of its argument which represents net profits. Keen and
subsequent authors recommend the following choice%
\begin{equation}
f\left( x\right) =p+\exp \left( qx+r\right) .  \label{Eq20a}
\end{equation}
\QTR{frametitle}{Solutions of KEs without Regularization are shown in Figure %
\ref{Fig4}.}

\begin{equation*}
\text{Figure \ref{Fig4} near here.}
\end{equation*}

On the one hand, these figures exhibit the desired features of the rapid
growth of firms' leverage. On the other hand, they produce an unrealistic
unemployment rate\textbf{\ }$\lambda _{w}>1$\textbf{.}

\subsection{Modified Theory}

{A simple modification along the lines outlined earlier, makes KEs more
credible:}%
\begin{eqnarray}
ds_{w} &=&-\left( a-b\lambda _{w}-\frac{\omega }{\lambda _{u}}\right)
s_{w}dt+\sigma _{s}\sqrt{s_{w}s_{f}}dW_{s}\left( t\right) ,  \label{Eq21} \\
d\lambda _{w} &=&\left( f\left( s_{f}-\frac{r_{L}\Gamma _{f}}{\nu _{f}}%
\right) -c-\frac{\omega }{s_{f}}\right) \lambda _{w}dt+\sigma _{\lambda }%
\sqrt{\lambda _{w}\lambda _{u}}dW_{\lambda }\left( t\right) ,  \notag \\
d\Gamma _{f} &=&\left( \left( r_{L}-\nu _{f}f\left( s_{f}-\frac{r_{L}\Gamma
_{f}}{\nu _{f}}\right) +d\right) \Gamma _{f}+\nu _{f}\left( f\left( s_{f}-%
\frac{r_{L}\Gamma _{f}}{\nu _{f}}\right) -s_{f}\right) \right) dt.  \notag
\end{eqnarray}%
Here $\omega $ is a regularization parameter, and $\sigma _{s}\sqrt{%
s_{w}s_{f}}$, $\sigma _{\lambda }\sqrt{\lambda _{w}\lambda _{u}}$ are Jacobi
normal volatilities.

\begin{sidewaysfigure}[t]
\center%
\includegraphics[width=1.0\textwidth, angle=0]{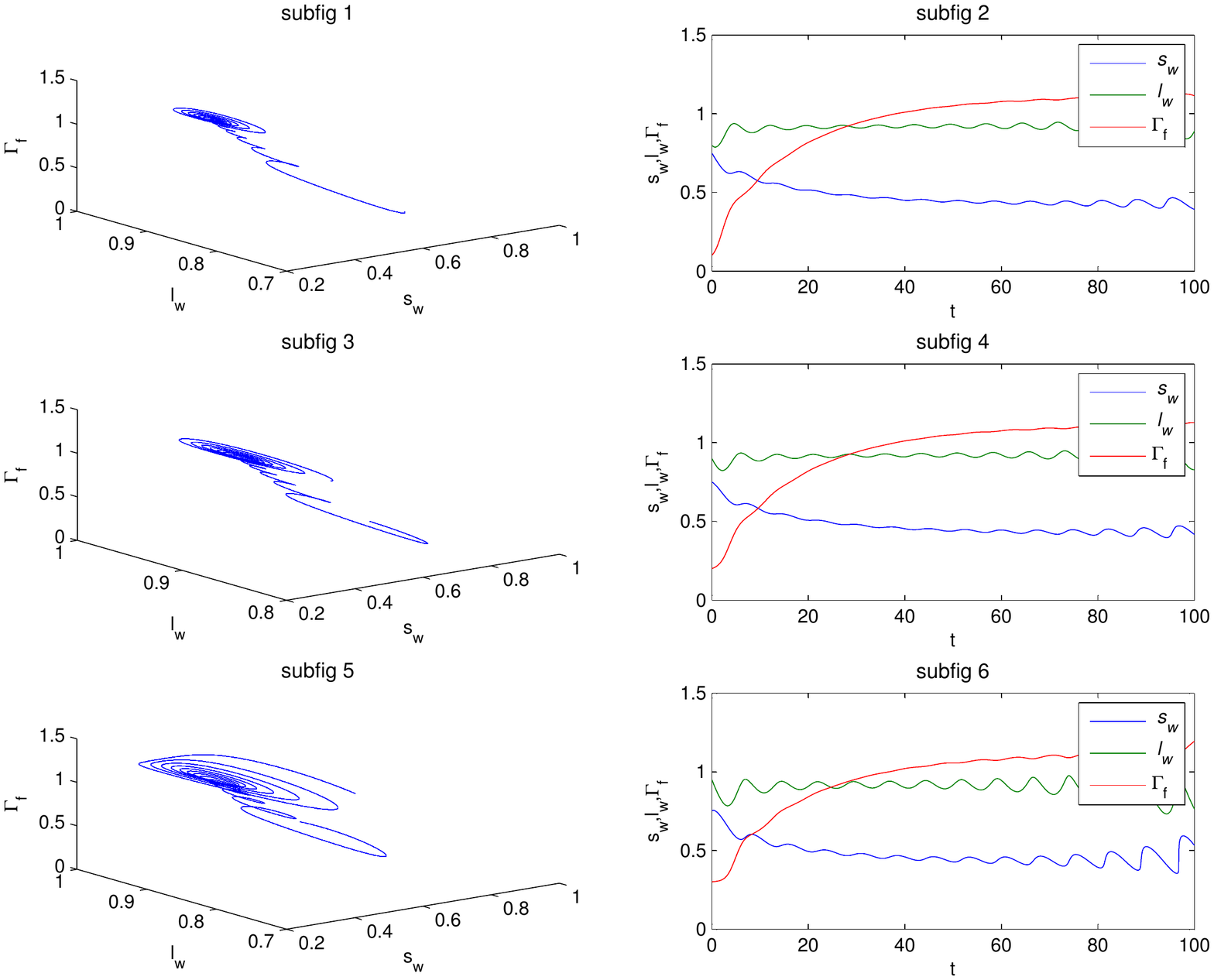}
\caption{{}Typical solutions of KEs with regularization.
Natural constraints are satisfied.  The same parameters and initial conditions as in Figure \ref{Fig4} are used;
in addition, $\omega=0.005$.}
\label{Fig5}
\end{sidewaysfigure}

\begin{sidewaysfigure}[t]
\center%
\includegraphics[width=1.0\textwidth, angle=0]{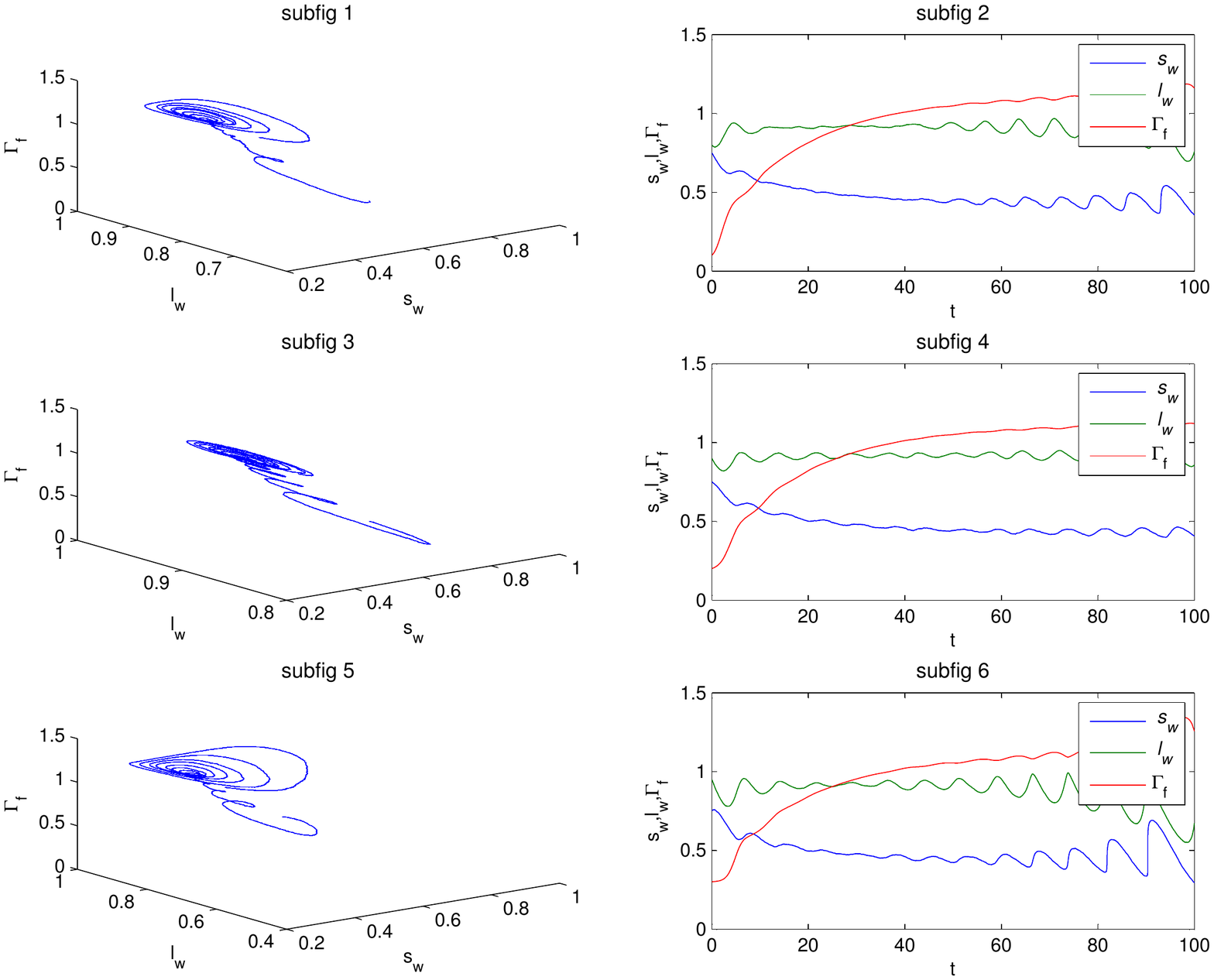}
\caption{{}Typical solutions of KEs with regularization and stochasticity.
Natural constraints are satisfied; orbits are non-periodic. 
The same parameters and initial conditions as in Figure \ref{Fig5} are used;
in addition, $\sigma_s=0.005, \sigma_l=0.005$.}
\label{Fig6}
\end{sidewaysfigure}

\QTR{frametitle}{Effects of regularization and effects of stochasticity
combined with regularization for KEs are presented in Figures \ref{Fig5}%
\thinspace\ and \ref{Fig6}, respectively.\footnote{\QTR{frametitle}{%
Partially regularized case without stochasticity is{\ also considered by
Grasselli \& Costa-Lima (2012).}}}}

\begin{equation*}
\text{Figure \ref{Fig5} near here.}
\end{equation*}

\begin{equation*}
\text{Figure \ref{Fig6} near here.}
\end{equation*}

While these Figures demonstrate the\textbf{\ }same\textbf{\ }rapid growth of
firms' leverage\textbf{\ }as in Figure\textbf{\ }\QTR{frametitle}{\ref{Fig4}}%
, while ensuring that $\lambda _{w}<1$, without taking into account a
possibility of defaults\textbf{\ }they are not detailed enough to describe
the approach of a crisis and the moment of the crisis itself.

Here and above we looked at the classical LVGEs and KEs and modified them to
better reflect the underlying economics. We use these equations as an
important building block for the stochastic MMC theory.

\section{A Simple Economy: Consumers, Producers, Banks\label{SimpleEconomy}}

\subsection{Inspiration}

Inspired by the above developments, we build a continuous-time stochastic%
\textbf{\ }model of the monetary circuit, which has attractive features of
the established models, but at the same time explicitly respects the fact
that production equals consumption plus investment, incorporate a
possibility of default by borrowers\textbf{, }satisfies all the relevant
economic constraints, and can be easily extended to integrate\textbf{\ }the
government and central bank, as well as other important aspects, in its
framework. For the first time, defaults by borrowers are\textbf{\ }%
explicitly incorporated into the model framework.

For the sake of brevity, we shall focus on a reduced monetary circuit
consisting of firms, banks, workers, and rentiers, while the extended
version will be reported elsewhere.

\subsection{Stocks and Flows}

\begin{figure}[t]
\center%
\includegraphics[width=1.0\textwidth, angle=0]{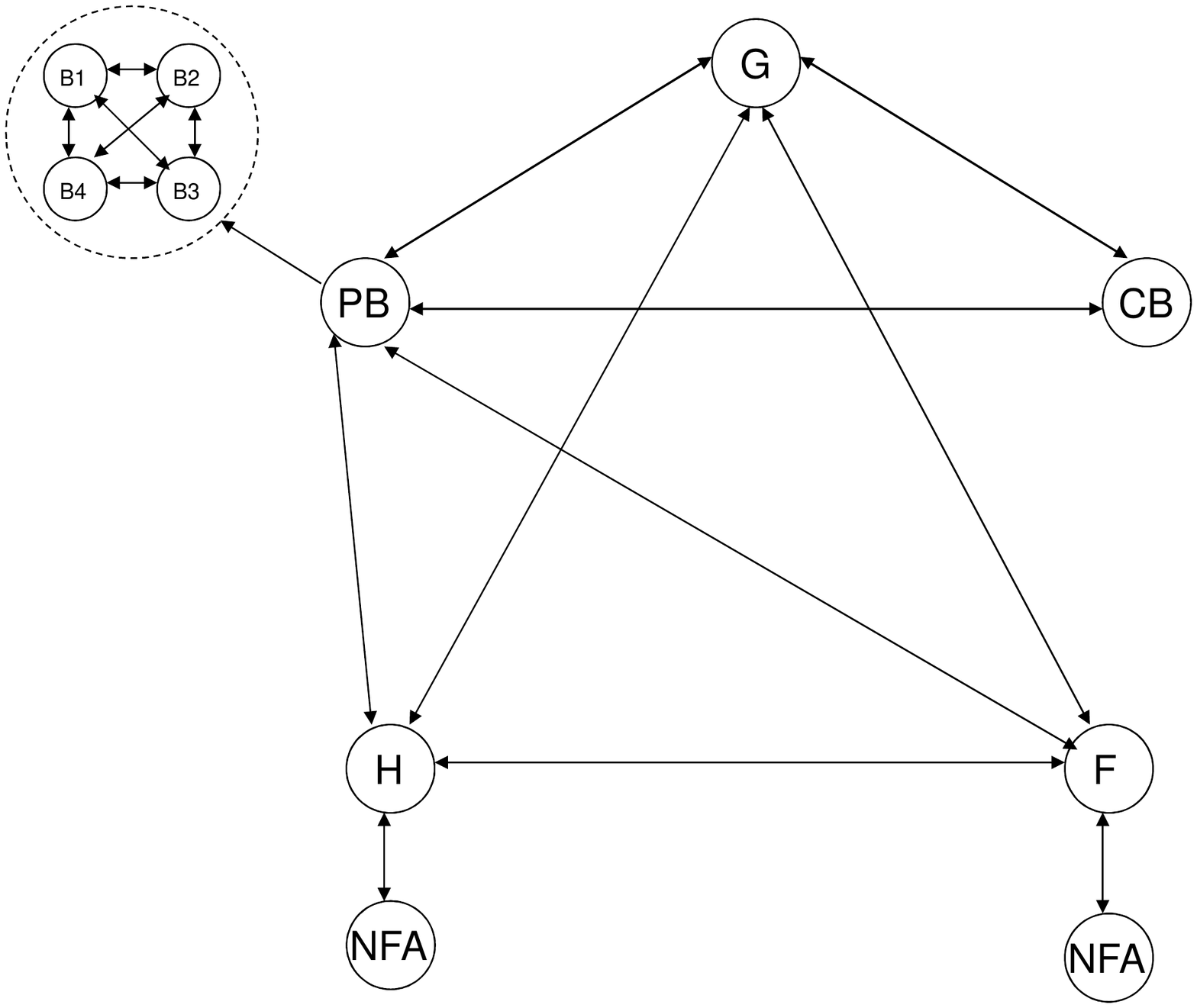}
\caption{{}Sketch of the Monetary Circuit. G - government, CB - central
bank, PB - private banks, F - firms, H - households including rentiers and
workers, NFA - non-financial assets.}
\label{Fig7}
\end{figure}

To describe the monetary circuit in detail, we need to consider five
sectors: households (workers and rentiers) $H$; firms (capitalists) $F$;
private banks (bankers) $PB$; government $G$; and central bank\textbf{\ }$CB$%
; all these sectors are presented in Figure \ref{Fig7}\ below. However, the
simplest viable economic graph with just three sectors, namely, households $%
H $\textbf{, }firms $F$\textbf{, }and private banks\textbf{\ }$PB$\textbf{,}
can produce a nontrivial monetary circuit, which is analyzed below\textbf{.}
Banks naturally play a central role in the monetary circuit by
simultaneously creating assets and liabilities. However, this crucial
function is performed under constraints on banks capital and liquidity.%
\textbf{\ }The emphasis on capital and liquidity in the general context of
monetary circuits is an important and novel feature, which differentiates
our approach from the existing ones.\textbf{\ }Further details, including
the role of the central bank as a\textbf{\ }system regulator, will be
reported elsewhere.

\begin{equation*}
\text{Figure \ref{Fig7} near here.}
\end{equation*}

\subsubsection{Notation}

We use subscripts $w,r,f,b$ to denote quantities related to workers,
rentiers, firms, and banks, respectively. We denote rentiers' and firms'
deposits (banks' liabilities) by $D_{r},D_{f}$, and their loans (banks'
assets) by $L_{r},L_{f}$. Firms' physical, non-financial assets are denoted
by $K_{f}$; banks' capital $K_{b}$; all these quantities are expressed in
monetary units, $\left[ M\right] $. Thus, financial and nonfinancial stocks
are denoted by $D_{r},L_{r},D_{f},L_{f},K_{f},K_{b}$. By its very nature,
bank capital, $K_{b}$ is a balancing variable between bank's assets $%
(L_{r}+L_{f})$ and liabilities $\left( D_{r}+D_{f}\right) $,%
\begin{equation}
K_{b}=L_{r}+L_{f}-\left( D_{r}+D_{f}\right) .  \label{Eq23}
\end{equation}%
According to banking regulations, bank assets are limited by the capital
constraints,%
\begin{equation}
K_{b}>\nu _{b}\left( L_{r}+L_{f}\right) ,  \label{Eq21a}
\end{equation}%
where $\nu _{b}$ is a non-dimensional capital adequacy ratio, which defines
the overall leverage in the financial system. When dealing with the banking
system as a whole, which, in essence can be viewed as a gigantic single
bank, we do not need to include the central bank, since the liquidity
squeeze cannot occur by definition. It goes without saying that when we deal
with a set of individual banks, the introduction of the central bank is an
absolute necessity. This extended case will be presented elsewhere.

There are several important rates, which determine monetary flows in our
simplified economy, namely, the deposit rate $r_{D}$, loan rate $r_{L}$,%
\footnote{%
We assume that $r_{D}$ is the same for rentiers and firms, and simplarly
with $r_{L}$.} maximum production rate at full employment, $v_{f}$, physical
assets amortization rate $\xi _{A}$, default rate $\xi _{\Delta }$; all
these rates are expressed in terms of inverse time units, $\left[ 1/T\right] 
$.

Contractual net interest cash flows for rentiers and firms, $ni_{r,f}$,
which are measured in terms of monetary units per time $\left[ M/T\right] $,
have the form%
\begin{equation}
ni_{r,f}=r_{D}D_{r,f}-r_{L}L_{r,f}.  \label{Eq22}
\end{equation}

Profits for firms and banks are denoted as $\Pi _{f}$ and $\Pi _{b}$,
respectively, with both quantities being expressed in monetary units per
time, $\left[ M/T\right] $.\textbf{\ }For future discussion, in addition to
the overall profits, we introduce distributed, $\Pi _{f}^{d}$\ and $\Pi
_{b}^{d}$, and undistributed, $\Pi _{f}^{u}$\ and $\Pi _{b}^{u}$, portions
of the profits.

It is also necessary to introduce various fractions, some of which we are
already familiar with, such as the workers' share of production $s_{w}$, the
firms' share of production $s_{f}=1-s_{w}$, employment rate $\lambda _{w}$,
unemployment rate $\lambda _{u}=1-\lambda _{w}$, and some of which are new,
such as capacity utilization $u_{f}$, the rentiers' share of firms' profits $%
\delta _{rf}$, the firms' share of the firms's profits $\delta
_{ff}=1-\delta _{rf}$, the rentiers' share of banks' profits $\delta _{rb}$,
the banks' share of the banks's profits $\delta _{bb}=1-\delta _{rb}$; all
these quantities are non-dimensional, $\left[ 1\right] ,$ and sandwiched
between 0 and 1. It is clear that $\Pi _{f}^{d}=\delta _{rf}\Pi _{f}$, etc.

\subsubsection{Key Observations}

(a) Production is equal to consumption plus investment:%
\begin{equation}
Y_{f}=C_{w}+C_{r}+I_{f}.  \label{Eq24}
\end{equation}%
All quantities in Eq. (\ref{Eq24}) are expressed in terms of $\left[ M/T%
\right] .$

(b) On the one hand, the workers' participation in the system is essentially
non-financial and amounts to straightforward exchange of labor for goods, so
that%
\begin{equation}
C_{w}=s_{w}Y_{f}.  \label{Eq25}
\end{equation}%
Thus, as was pointed out by Kaletcki, workers consume what they earn.

(c) On the other hand, rentiers\textbf{\ }can discretionally choose their
level of consumption\textbf{, }$C_{r},$\textbf{\ }introducing therefore the
notion of stochasticity into the picture. We explicitly model the stochastic
nature of their consumption by assuming that it is governed by the SDE of
the form%
\begin{eqnarray}
dC_{r} &=&\kappa \left( \bar{C}_{r}-C_{r}\right) dt+\sigma C_{r}dW_{C}\left(
t\right)  \label{Eq26} \\
\bar{C}_{r} &=&\alpha _{0}\left( ni_{r}+\Pi _{f}^{d}+\Pi _{b}^{d}\right)
+\alpha _{1}\nu _{f}K_{f},  \notag
\end{eqnarray}%
where we use the fact that total stock $\Sigma _{r}$ of financial and
non-financial assets belonging to rentiers (as a class) is given by%
\begin{eqnarray}
\Sigma _{r} &=&D_{r}-L_{r}+K_{f}+D_{f}-L_{f}+K_{b}  \label{Eq27} \\
&=&D_{r}-L_{r}+K_{f}+D_{f}-L_{f}+L_{r}+L_{f}-D_{r}-D_{f}  \notag \\
&=&K_{f}.  \notag
\end{eqnarray}%
In other words, the rentiers' property boils down to firms' non-financial
assets. Eqs (\ref{Eq26}) assume that rentiers' consumption is reverting to
the mean, $\bar{C}_{r},$ which is a linear combination of profits received
by rentiers\textbf{, }$ni_{r}+\Pi _{f}^{d}+\Pi _{b}^{d}$\textbf{, }and the
theoretical productivity of their capital, $\nu _{f}K_{f}$.

(d) We apply the celebrated Hooke's law and assume that firms invest in
proportion to their overall production%
\begin{equation}
I_{f}=\gamma _{f}Y_{f}  \label{Eq28}
\end{equation}%
We view this law\textbf{\ }as a first-order linearization of any
hyperelastic relation, which exists in practice. Thus, firms' production
depends on rentiers' consumption%
\begin{equation}
Y_{f}=\frac{C_{r}}{s_{f}-\gamma _{f}},  \label{Eq29}
\end{equation}%
Here we assume that firms reinvest in production out of the share of their
profits, so that $0<\gamma _{f}<s_{f}$, keeping $C_{r}$ positive, $C_{r}>0$.
It is convenient to represent $\gamma _{f}$ in the form%
\begin{equation}
\gamma _{f}=\upsilon _{f}s_{f},  \label{Eq30}
\end{equation}%
$0<\upsilon _{f}<1$, and represent $Y_{f}$ in the form%
\begin{equation}
Y_{f}=\frac{C_{r}}{\left( 1-\upsilon _{f}\right) s_{f}}.  \label{Eq31}
\end{equation}%
(e) Thus, the level of investment and capacity utilization are given by%
\begin{equation}
I_{f}=\frac{\upsilon _{f}C_{r}}{\left( 1-\upsilon _{f}\right) },
\label{Eq32}
\end{equation}%
\begin{equation}
u_{f}=\frac{Y_{f}}{\nu _{f}K_{f}}=\frac{C_{r}}{\left( s_{f}-\gamma
_{f}\right) \nu _{f}K_{f}}=\frac{C_{r}}{\left( 1-\upsilon _{f}\right)
s_{f}\nu _{f}K_{f}}.  \label{Eq33}
\end{equation}%
(f) Firms' overall profits, distributed, and undistributed, are defined as%
\begin{eqnarray}
\Pi _{f} &=&s_{f}Y_{f}+r_{D}D_{f}-r_{L}L_{f}=\frac{C_{r}}{\left( 1-\upsilon
_{f}\right) }+ni_{f},  \label{Eq34} \\
\Pi _{f}^{d} &=&\delta _{rf}\Pi _{f},\ \ \ \ \ \Pi _{f}^{u}=\delta _{ff}\Pi
_{f}.  \notag
\end{eqnarray}%
Thus, firms' profits are directly proportional to rentiers consumption. As
usual, Kaletcki put it best by observing that capitalists earn what they
spend!

The dimensionless profit rate $\pi _{f}$ is%
\begin{equation}
\pi _{f}=\frac{\Pi _{f}}{K_{f}}.  \label{Eq35}
\end{equation}%
The proportionality coefficient $\upsilon _{f}$ introduced in Eq.\ (\ref%
{Eq30}) depends on the profit rate, capacity utilization, financial
leverage, etc., so that%
\begin{equation}
\upsilon _{f}=\Phi \left( \upsilon _{0}+\upsilon _{1}\frac{s_{f}Y_{f}}{\nu
_{f}K_{f}}+\upsilon _{2}\frac{D_{f}}{K_{f}}+\upsilon _{3}\frac{L_{f}}{K_{f}}%
\right) ,  \label{Eq36}
\end{equation}%
or, explicitly,%
\begin{equation}
\upsilon _{f}=\Phi \left( \upsilon _{0}+\upsilon _{1}\frac{C_{r}}{\left(
1-\upsilon _{f}\right) \nu _{f}K_{f}}+\upsilon _{2}\frac{D_{f}}{K_{f}}%
+\upsilon _{3}\frac{L_{f}}{K_{f}}\right) ,  \label{Eq37}
\end{equation}%
where $\Phi \left( .\right) $ maps the real axis onto the unit interval,
constants $\upsilon _{0},\upsilon _{1},\upsilon _{2}$ are positive, and
constant $\upsilon _{3}$ is negative. We choose $\Phi $ in the form 
\begin{equation}
\Phi \left( x\right) =\frac{1}{1+\exp \left( -2x\right) }.  \label{Eq37a}
\end{equation}

(g) Banks' overall profits,\textbf{\ }distributed, and undistributed,\textbf{%
\ }represent the difference between interest received on outstanding loans
and paid on banks deposits reduced by defaults on loans, so that%
\begin{equation}
\Pi _{b}=-\xi _{\Delta }\left( L_{r}+L_{f}\right) -ni_{r}-ni_{f},\ \ \ \ \
\Pi _{b}^{d}=\delta _{rb}\Pi _{b},\ \ \ \ \ \Pi _{b}^{u}=\delta _{bb}\Pi
_{b}.  \label{Eq38}
\end{equation}

(h) Rentiers' cash flows are%
\begin{equation}
CF_{r}=r_{D}D_{r}-r_{L}L_{r}+\ \Pi _{f}^{d}+\ \Pi _{b}^{d}-C_{r}=ni_{r}+\
\Pi _{f}^{d}+\ \Pi _{b}^{d}-C_{r}.  \label{Eq39}
\end{equation}%
If $CF_{r}>0$, then rentiers'\textbf{\ }deposits\textbf{, }$D_{r}$,
increase, otherwise, their loans, $L_{r}$, increase. Thus%
\begin{equation}
dD_{r}=\left( ni_{r}+\ \Pi _{f}^{d}+\ \Pi _{b}^{d}-C_{r}\right)
^{+}dt=\left( CF_{r}\right) ^{+}dt,  \label{Eq40}
\end{equation}%
\begin{eqnarray}
dL_{r} &=&-\xi _{\Delta }L_{r}dt+\left( -ni_{r}-\ \Pi _{f}^{d}-\ \Pi
_{b}^{d}+C_{r}\right) ^{+}dt  \label{Eq41} \\
&=&-\xi _{\Delta }L_{r}dt+\left( -CF_{r}\right) ^{+}dt.  \notag
\end{eqnarray}%
This equation takes into account a possibility of rentiers' default.

(i) Firms' cash flows\textbf{\ }are%
\begin{equation}
CF_{f}=\Pi _{f}^{u}-\gamma _{f}Y_{f}.  \label{Eq42}
\end{equation}%
If $CF_{f}>0$\textbf{, }then firms'\textbf{\ }deposits\textbf{, }$D_{f}$,
increase, otherwise, their loans, $L_{f}$, increase. Thus%
\begin{equation}
dD_{f}=\left( \Pi _{f}^{u}-\gamma _{f}Y_{f}\right) ^{+}dt=\left(
CF_{f}\right) ^{+}dt,  \label{Eq43}
\end{equation}

\begin{eqnarray}
dL_{f} &=&-\xi _{\Delta }L_{f}dt+\left( -\Pi _{f}^{u}+\gamma
_{f}Y_{f}\right) ^{-}dt  \label{Eq44} \\
&=&-\xi _{\Delta }L_{f}dt+\left( -CF_{f}\right) ^{+}dt.  \notag
\end{eqnarray}%
The latter equation takes into account a possibility of firms' default.

(j) Firms' physical assets growth depends on their investments and the rate
of depreciation,\textbf{\ }%
\begin{equation}
dK_{f}=\left( \frac{\upsilon _{f}C_{r}}{1-\upsilon _{f}}-\xi
_{A}K_{f}\right) dt.  \label{Eq45}
\end{equation}%
\qquad

(k) Banks' capital growth is determined by their net interest income and the
rate of default,\textbf{\ }%
\begin{equation}
dK_{b}=\Pi _{b}^{u}dt=\delta _{bb}\left( -\xi _{\Delta }\left(
L_{r}+L_{f}\right) -ni_{r}-ni_{f}\right) .  \label{Eq46}
\end{equation}

(l) Physical and Financial capacity constraints (at full employment)\textbf{%
\ }have the form%
\begin{equation}
Y_{f}=\min \left( Y_{f},\nu _{f}K_{f}\right) ,  \label{Eq47}
\end{equation}%
\begin{equation}
\left( -CF_{b}\right) ^{+}=\left( -CF_{b}\right) ^{+}\mathbb{I}_{\nu
_{b}\left( L_{r}+L_{f}\right) -K_{b}<0},  \label{Eq48}
\end{equation}%
\begin{equation}
\left( -CF_{f}\right) ^{+}=\left( -CF_{f}\right) ^{+}\mathbb{I}_{\nu
_{b}\left( L_{r}+L_{f}\right) -K_{b}<0}.  \label{Eq49}
\end{equation}%
We emphasize this direct parallel between financial and non-financial
worlds, with\ the capital ratio playing the role of a physical capacity
constraint.

(m) We use the above observations to derive a modified version of the LVGEs (%
\ref{Eq16}). While the first equation describing the dynamics for $s_{w}$\
remains unchanged, the second equation for $\lambda _{w}$ becomes%
\begin{eqnarray}
d\lambda _{w} &=&\left( \frac{I_{f}}{\nu _{f}K_{f}}-\alpha -\beta -\xi
_{A}\right) \lambda _{w}dt  \label{Eq52} \\
&=&\left( \frac{\upsilon _{f}}{\left( 1-\upsilon _{f}\right) }\frac{C_{r}}{%
\nu _{f}K_{f}}-\alpha -\beta -\xi _{A}\right) \lambda _{w}dt,  \notag
\end{eqnarray}%
or, symbolically,%
\begin{equation}
d\lambda _{w}=\left( \frac{\upsilon _{f}}{\left( 1-\upsilon _{f}\right) }%
\frac{C_{r}}{\nu _{f}K_{f}}-c\right) \lambda _{w}dt.  \label{Eq52a}
\end{equation}

(n) By using Eqs (\ref{Eq4}) and (\ref{Eq31}), we can express the level of
prices, $P$, as a function of rentiers' consumption, $C_{r}$, employment, $%
\lambda _{w}$, and other important economic variables. These equations show
that%
\begin{equation}
\frac{C_{r}}{\left( 1-\upsilon _{f}\right) s_{f}}=\lambda _{w}\theta
_{w}N_{w}P.  \label{Eq50}
\end{equation}%
Accordingly, we can represent $P$ as follows%
\begin{equation}
P=\frac{C_{r}}{\left( 1-\upsilon _{f}\right) s_{f}\lambda _{w}\theta
_{w}N_{w}}.  \label{Eq51}
\end{equation}

\subsection{Main Equations}

In this section we summarize the main dynamic MMC equations and the
corresponding constraints%
\begin{equation}
\begin{array}{l}
dC_{r}=\kappa _{C}\left( \bar{C}_{r}-C_{r}\right) dt+\sigma
_{C}C_{r}dW_{C}\left( t\right) , \\ 
dD_{r}=\left( \delta _{bb}ni_{r}+\left( \delta _{rf}-\delta _{rb}\right)
ni_{f}-\delta _{rb}\xi _{\Delta }\left( L_{r}+L_{f}\right) -\frac{\left(
\delta _{ff}-\upsilon _{f}\right) C_{r}}{\left( 1-\upsilon _{f}\right) }%
\right) ^{+}dt, \\ 
dL_{r}=\left( -\xi _{\Delta }L_{r}+\left( -\delta _{bb}ni_{r}-\left( \delta
_{rf}-\delta _{rb}\right) ni_{f}+\delta _{rb}\xi _{\Delta }\left(
L_{r}+L_{f}\right) +\frac{\left( \delta _{ff}-\upsilon _{f}\right) C_{r}}{%
\left( 1-\upsilon _{f}\right) }\right) ^{+}\right) dt, \\ 
dD_{f}=\left( \delta _{ff}ni_{f}+\frac{\left( \delta _{ff}-\upsilon
_{f}\right) C_{r}}{\left( 1-\upsilon _{f}\right) }\right) ^{+}dt, \\ 
dL_{f}=\left( -\xi _{\Delta }L_{f}+\left( -\delta _{ff}ni_{f}-\frac{\left(
\delta _{ff}-\upsilon _{f}\right) C_{r}}{\left( 1-\upsilon _{f}\right) }%
\right) ^{+}\right) dt, \\ 
dK_{f}=\left( \frac{\upsilon _{f}C_{r}}{\left( 1-\upsilon _{f}\right) }-\xi
_{A}K_{f}\right) dt+\sigma _{K}K_{f}dW_{K}\left( t\right) , \\ 
dK_{b}=-\delta _{bb}\left( \xi _{\Delta }\left( L_{r}+L_{f}\right)
+ni_{r}+ni_{f}\right) ,%
\end{array}
\label{Eq53}
\end{equation}%
where%
\begin{equation}
\begin{array}{l}
ni_{r,f}=r_{D}D_{r,f}-r_{L}L_{r,f}, \\ 
\ \ \bar{C}_{r}=\alpha _{0}\left( \delta _{bb}ni_{r}+\left( \delta
_{rf}-\delta _{rb}\right) ni_{f}+\frac{\delta _{rf}C_{r}}{\left( 1-\upsilon
_{f}\right) }\right) +\alpha _{1}\nu _{f}K_{f}, \\ 
\ \ \ \upsilon _{f}=\Phi \left( \upsilon _{0}+\upsilon _{1}\frac{C_{r}}{%
\left( 1-\upsilon _{f}\right) \nu _{f}K_{f}}+\upsilon _{2}\frac{D_{f}}{K_{f}}%
+\upsilon _{3}\frac{L_{f}}{K_{f}}\right) .%
\end{array}
\label{Eq54}
\end{equation}%
The coefficient $\upsilon _{f}$\ introduced in Eq. (\ref{Eq30}) can be found
either via the Newton-Raphson method or via fixed-point iteration. The first
iteration is generally sufficient, so that, approximately,%
\begin{equation}
\upsilon _{f}\approx \Phi \left( \upsilon _{0}+\upsilon _{1}\frac{C_{r}}{%
\left( 1-\Phi \left( \upsilon _{0}\right) \right) \nu _{f}K_{f}}+\upsilon
_{2}\frac{D_{f}}{K_{f}}+\upsilon _{3}\frac{L_{f}}{K_{f}}\right) .
\label{Eq55}
\end{equation}

The physical and financial capacity constraints are%
\begin{equation}
\begin{array}{l}
\ \ \ \ \ \ \ \ \ \ \ Y_{f}=\min \left( Y_{f},\nu _{f}K_{f}\right) , \\ 
\left( -CF_{b}\right) ^{+}=\left( -CF_{b}\right) ^{+}\mathbb{I}_{\nu
_{b}\left( L_{r}+L_{f}\right) -K_{b}<0}, \\ 
\left( -CF_{f}\right) ^{+}=\left( -CF_{f}\right) ^{+}\mathbb{I}_{\nu
_{b}\left( L_{r}+L_{f}\right) -K_{b}<0}.%
\end{array}
\label{Eq55a}
\end{equation}

In addition,%
\begin{equation}
\begin{array}{l}
\ d\theta _{w}=\alpha \theta _{w}dt, \\ 
dN_{w}=\beta N_{w}dt, \\ 
\ ds_{w}=-\left( a-b\lambda _{w}-\frac{\omega }{\lambda _{u}}\right)
s_{w}dt+\sigma _{s}\sqrt{s_{w}s_{f}}dW_{s}\left( t\right) , \\ 
d\lambda _{w}=\left( \frac{\upsilon _{f}C_{r}}{\left( 1-\upsilon _{f}\right)
\nu _{f}K_{f}}-c-\frac{\omega }{s_{f}}\right) \lambda _{w}dt+\sigma
_{\lambda }\sqrt{\lambda _{w}\lambda _{u}}dW_{\lambda }\left( t\right) , \\ 
\ \ \ P=\frac{C_{r}}{\left( 1-\upsilon _{f}\right) s_{f}\lambda _{w}\theta
_{w}N_{w}}.%
\end{array}
\label{Eq56}
\end{equation}

In summary, we propose the closed system of stochastic scale invariant MMC
equations (\ref{Eq53}), (\ref{Eq54}). By construction, these equation
preserve the equality among production and consumption plus investment. In
addition, it turns out modified LVGEs play only an auxiliary role and are
not necessary for understanding the monetary circuit at the most basic
level. This intriguing property is due to the assumption that investments as
driven solely by profits. If capacity utilization is incorporated into the
picture, then MMC equations and LVGEs become interlinked.

\begin{sidewaysfigure}[t]
\center%
\includegraphics[width=0.75\textwidth, angle=0]{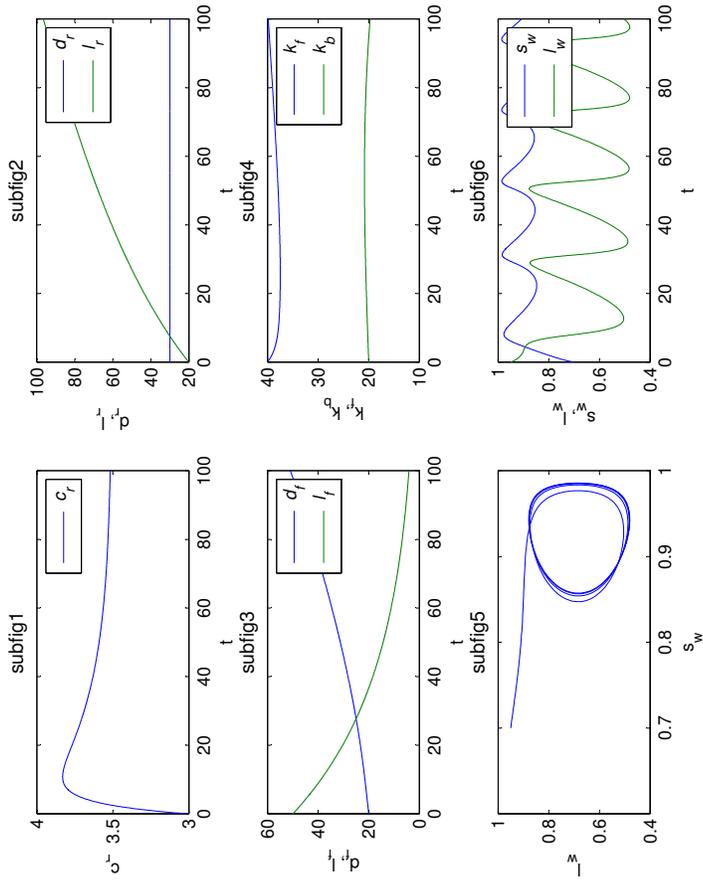}
\caption{{}Evolution of various quantities of interest in the non-stochastic circuit. Representative parameters and initial conditions are as follows:
$\kappa_C=0.5, \delta_bb=0.5, \delta_rb=0.5, \delta_ff=0.25, \delta_rf=0.75, \xi_\Delta=0.025, \xi_A=0.02, r_D=0.02, 
r_L=0.04, \nu=0.13, \alpha_0=0.5, \alpha_1=0.5, \upsilon_0=-1.6, \upsilon_1=1.1, \upsilon_2=0.1, \upsilon_0=-0.2,
 a=0.05, b=0.05, c=0.075, \omega=0.005$, $C_r=3, D_r=30, L_r=20, D_f=20, L_f=50, K_f=40, K_b=20, s_w=0.7, \lambda_w=0.95$.}
\label{Fig8}
\end{sidewaysfigure}

Representative solution of MMC equations is shown in Figure \ref{Fig8}.

\begin{equation*}
\text{Figure \ref{Fig8} near here.}
\end{equation*}

\section{Money Creation and Annihilation in Pictures\label{MoneyCreation}}

In modern society, where large quantities of money have to be deposited in
banks, banks play a unique role as record keepers.\footnote{%
In general, in developed economies the proportion of cash versus bank
deposits is rather small. However, when very large denomination notes are
available, they are frequently used in lieu of bank accounts.} Depositors
become, in effect, unsecured junior creditors of banks. If a bank were to
default, it would generally cause partial destruction of deposits. To avoid
such a disturbing eventuality, banks are required to keep sufficient capital
cushions, as well as ample liquidity. In addition, deposits are insured up
to a certain threshold. Without diving into nuances of different takes on
the nature of banking\textbf{,} we mention several books and papers written
over the last century, which reflect upon various pertinent issues, such as
Schumpeter (1912), Howe (1915), Klein (1971), Saving (1977), Sealey and
Lindley (1977), Diamond and Dybvig (1983), Fama (1985), Selgin and White
(1987), Heffernan (1996), FRB (2005), Wolf (2014).

\begin{sidewaysfigure}[t]
\centering%
\fbox{\includegraphics[width = 1.0\textwidth]{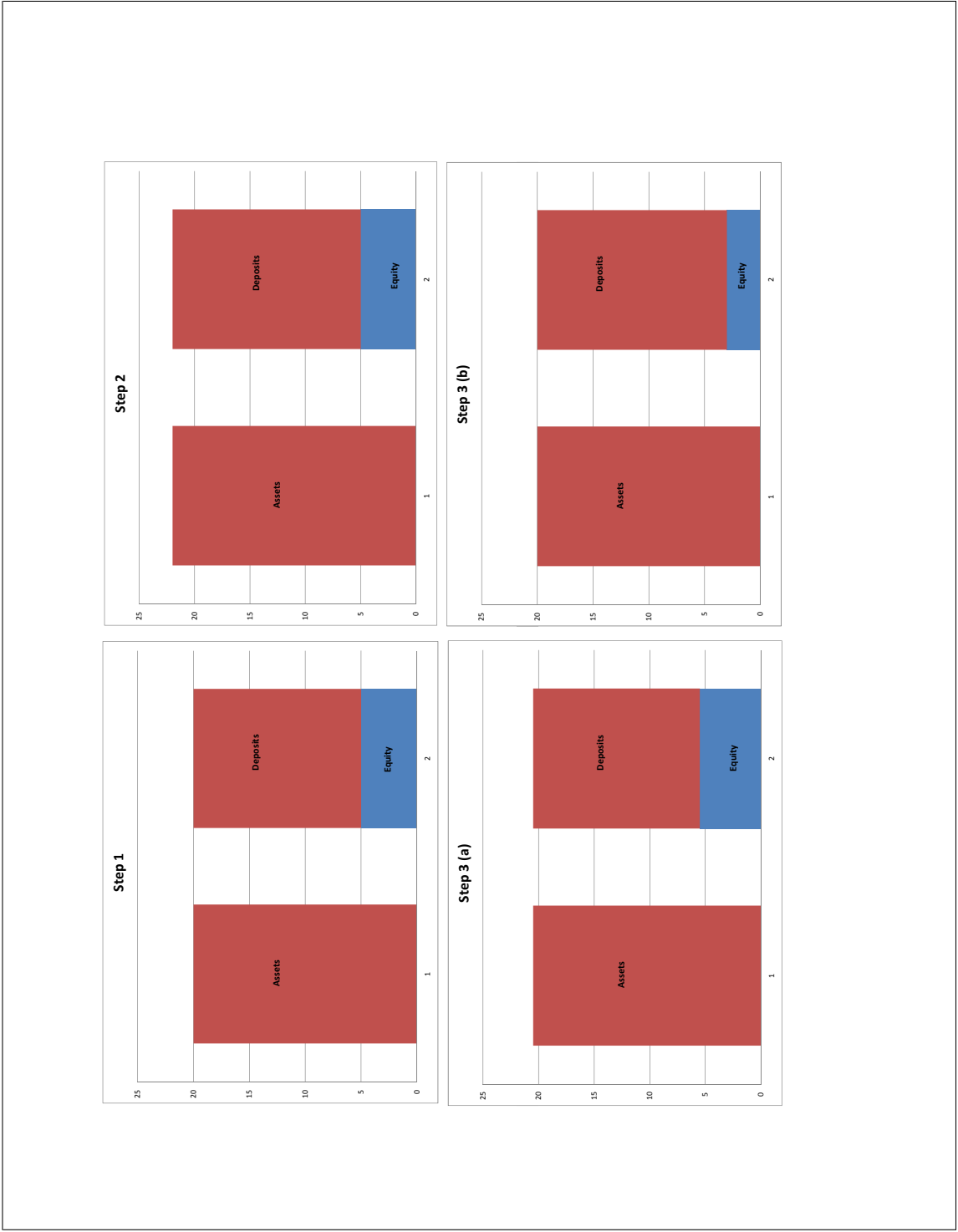}}
\caption{Credit money creation and annihilation by one commercial bank.}
\label{Fig9}
\end{sidewaysfigure}

It is very useful to have a simplified pictorial representation for the
inner working of the banking system. We start with a simple case of a single
bank, or, equivalently, the banking system as a whole. We assume that the
bank in question does not operate at full capacity, so that condition (\ref%
{Eq21a}) is satisfied. If a new borrower\textbf{, }who is deemed to be
credit worthy, approaches the bank and asks for a reasonably-sized loan,
then the bank issues the loan by simultaneously creating on its books a
deposit (the borrower's asset)\textbf{, }and a matching liability for the
borrower (the bank's asset). Figuratively speaking, the bank has created%
\textbf{\ }money "out of thin air". Of course, when the loan is repaid, the
process is carried in reverse, and the money is \textbf{"}destroyed\textbf{"}%
. Assuming that the interest charged on loans is greater than the interest
paid on deposits, as a result of the round-trip process bank's capital
increases.\footnote{%
The money is destroyed if the borrower defaults, as well. It this case,
however, bank's capital naturally decreases.} The whole process, which is
relatively simple, is illustrated in Figure \ref{Fig9}. At first, the bank
has 20 units of assets, 15 units of liabilities, and 5 units of equity.
Then, it lends 2 units to a credit worthy borrower. Now it has 22 units of
assets and 17 units of liabilities. Thus, 2 units of new money are created.
If the borrower repays her debt with interest, as shown in Step 3(a), then
the bank accumulates 20.5 units of assets, 15 units of liabilities, and 5.5
units of equity. If the borrower defaults, as shown in Step 3(b), then the
bank ends up with 20 units of assets, 17 units of liabilities, and 3 units
of equity. In both cases 2 units of money are destroyed.

\begin{equation*}
\text{Figure \ref{Fig9} near here.}
\end{equation*}

Werner executed this process step by step and described his experiences in a
recent paper (Werner 2014). It is worth noting, that in the case of a single
bank, lending activity is limited by bank's capital capacity only and
liquidity is not important.

We now consider a more complicated case of two (or, possibly, more) banks.
In this case, it is necessary to incorporate liquidity into the picture. To
this end, we also must include a central bank into the financial ecosystem.
We assume that banks keep part of their assets in cash, which represents a
liability of the central bank.\footnote{%
Here cash is understood as an electronic record in the central bank ledger.}
The money creation process comprises of three stages: (a) A credit worthy
borrower asks the first bank for a loan, which the bank grants out of its
cash reserves, thus reducing its liquidity below the desired level; (b) The
borrower then deposits the money with the second bank, which converts this
deposit into cash, thereby increasing\textbf{\ }its liquidity above its
desired level\textbf{;} (c) The\textbf{\ }first bank approaches the second
bank in order to borrow its excess cash. If the second bank deems the first
bank credit worthy, it will lend its excess cash, in consequence creating 
\textbf{\ }a link between itself and the first bank. Alternatively, if the
second bank refuses to lend its excess cash to the first bank\textbf{, }the
first bank has to borrow funds from the central bank, by using its
performing assets as collateral. Thus, the central bank lubricates the
wheels of commerce by providing liquidity to\textbf{\ }credit worthy
borrowers. Its willingness to lend money to commercial banks, determines in
turn their willingness to lend to firms and households. When the borrower
repays its loan the process plays in reverse.

\begin{sidewaysfigure}[t]
\centering%
\fbox{\includegraphics[width = 1.0\textwidth]{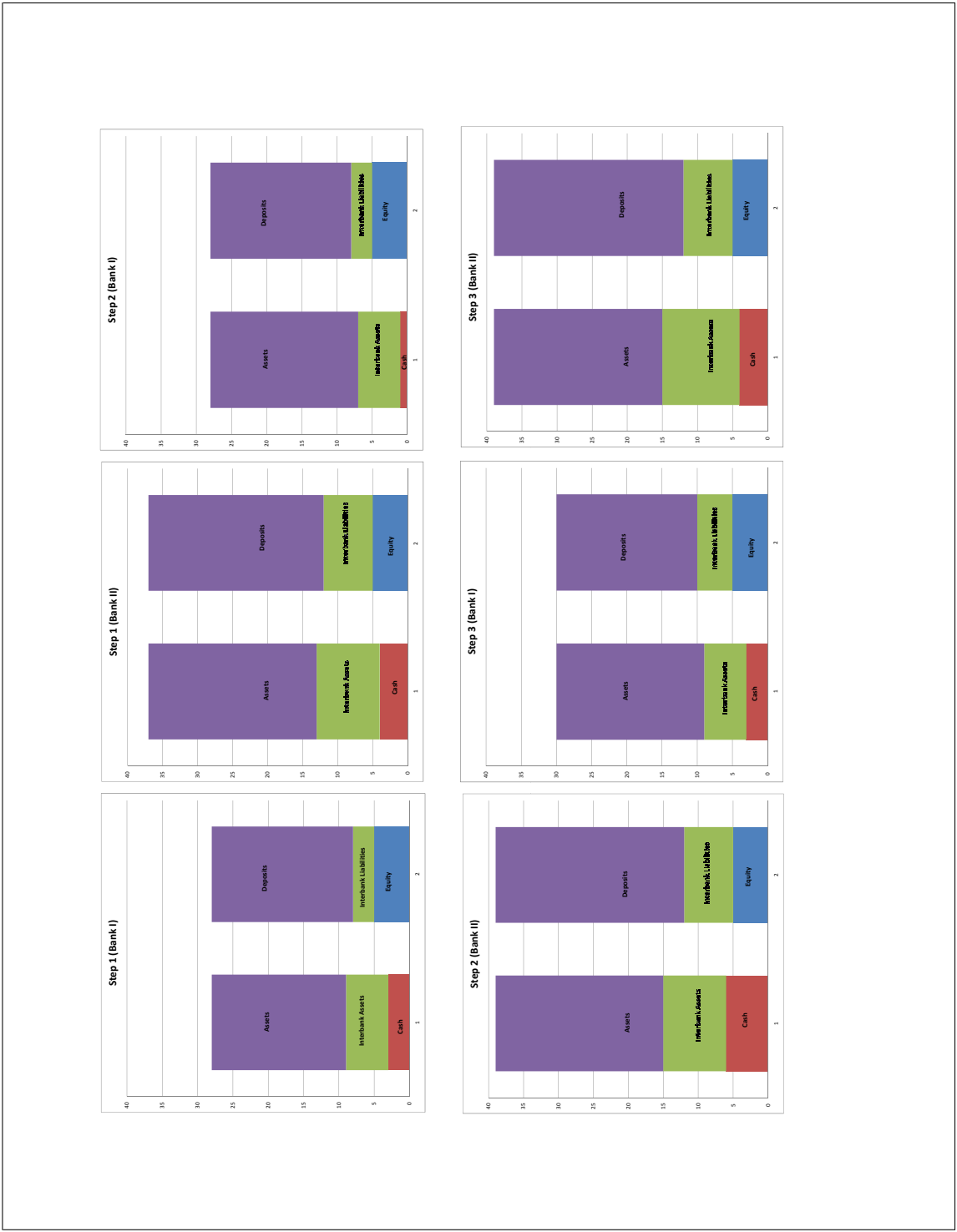}}
\caption{Credit money creation by two commercial banks.}
\label{Fig10}
\end{sidewaysfigure}

The money creation process, initiated when Bank I lends 2 monetary units to
a new borrower, results in the following changes in two banks' balance
sheets:%
\begin{equation}
\begin{array}{c}
\\ 
\\ 
\text{External\ Assets} \\ 
\text{Interbank\ Assets} \\ 
\text{Cash} \\ 
\text{External\ Liabilities} \\ 
\text{Interbank\ Liabilities} \\ 
\text{Equity}%
\end{array}%
\begin{array}{cc}
\text{Step} & \text{I} \\ 
\text{Bank\ I} & \text{Bank\ II} \\ 
19 & 24 \\ 
6 & 9 \\ 
3 & 4 \\ 
20 & 25 \\ 
3 & 7 \\ 
5 & 5%
\end{array}%
\begin{array}{cc}
\text{Step} & \text{II} \\ 
\text{Bank\ I} & \text{Bank\ II} \\ 
21 & 24 \\ 
6 & 9 \\ 
1 & 6 \\ 
20 & 27 \\ 
3 & 7 \\ 
5 & 5%
\end{array}%
\begin{array}{cc}
\text{Step} & \text{III} \\ 
\text{Bank\ I} & \text{Bank\ II} \\ 
21 & 24 \\ 
6 & 11 \\ 
3 & 4 \\ 
20 & 27 \\ 
5 & 7 \\ 
5 & 5%
\end{array}
\label{Eq56b}
\end{equation}%
This process is illustrated in Figure \ref{Fig10}. We leave it to the reader
to analyze the money annihilation process.

\begin{equation*}
\text{Figure \ref{Fig10} near here.}
\end{equation*}

\begin{sidewaysfigure}[t]
\centering%
\makebox[\textwidth]{
  \subfigure[]
  {
  	\includegraphics[width = 0.5\textwidth]{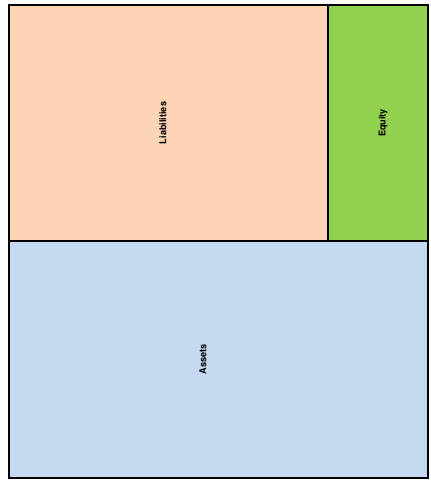}
  }
  \subfigure[]
  {
  	\includegraphics[width = 0.5\textwidth]{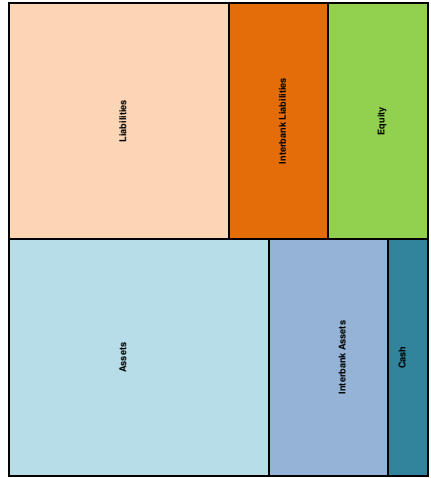}
  }
}
\caption{Comparison of balance sheets of a non-financial company (subfigure a), 
and a representative commercial bank (subfigure b).}
\label{Fig11}
\end{sidewaysfigure}

In summary, in contrast to a non-banking firm, whose balance sheet can be
adequately described by a simple relationship among assets, $A$,
liabilities, $L$, and equity, $E$,%
\begin{equation}
A=L+E,  \label{Eq56a}
\end{equation}%
as is shown in Figure \ref{Fig11}a, the balance sheet of a typical
commercial bank must, in addition to external assets and liabilities,
incorporate more details, such as interbank assets and liabilities, as well
as cash, representing simultaneously bank's assets and central bank's
liabilities, see Figure \ref{Fig11}b.

\begin{equation*}
\text{Figure \ref{Fig11} near here.}
\end{equation*}

In Section \ref{SimpleEconomy} we quantitatively described a supply and
demand driven economic system. In this system money is treated on a par with
other goods, and the dynamics of demand for loans and lending activity is
understood in the supply-demand equilibrium framework. An increasing demand
for loans from firms and households leads banks to lend more. Having said
that, we should emphasize that the ability of banks to generate new loans is
not infinite. In exact parallel with physical goods, whose overall
production at full employment is physically limited by $\nu _{f}K_{f}$, the
process of money (loan) creation is limited by the capital capacity of the
banking system $K_{b}/\nu _{b}$.\textbf{\ }Once we have embedded the flow of
money in the supply-demand framework, we can extend the model to several
interconnected banks that issue loans in the economy. These banks compete
with each other for business, while, at the same time, help each other to
balance their cash holdings thus creating interbank linkages. These linkages
are posing risks because of potential propagation of defaults in the system.%
\textbf{\ }Our main goal in the next section is to develop a parsimonious
model which, nevertheless, is rich enough to produce an adequate
quantitative description of the banking ecosystem. We look for a model with
as few adjustable parameters as possible rather than one over-fitted with a
plethora of adjustable calibration parameters.

\section{Interlinked Banking System\label{BankingSystem}}

Consider $N$ banks with external as well as mutual assets and liabilities of
the form%
\begin{equation}
A_{i}+\dsum\limits_{j\neq i}L_{ji}=A_{i}+\hat{A}_{i}\text{ and }%
L_{i}+\dsum\limits_{j\neq i}L_{ij}=L_{i}+\hat{L}_{i},\ \ i,j=1,...,N,
\label{Eq57}
\end{equation}%
where the interbank assets and liabilities are defined as%
\begin{equation}
\hat{A}_{i}=\dsum\limits_{j\neq i}L_{ji},\ \ \ \ \ \hat{L}%
_{i}=\dsum\limits_{j\neq i}L_{ij}.  \label{Eq59}
\end{equation}%
Accordingly, an individual bank's capital is given by%
\begin{equation}
E_{i}=A_{i}+\hat{A}_{i}-L_{i}-\hat{L}_{i}.  \label{Eq60}
\end{equation}%
We can represent banks assets and liabilities by using the following asset
and liability matrices%
\begin{equation}
\begin{array}{c}
\mathbb{A}=\left( A_{ij}\right) ,\ \ \ A_{ii}=A_{i},\ \ \ A_{ij}=L_{ji}, \\ 
\mathbb{L=}\left( L_{ij}\right) ,\ \ \ L_{ii}\equiv L_{i},\ \ \ i,j=1,...,N.%
\end{array}
\label{Eq61}
\end{equation}

Thus, by its very nature banking system becomes inherently linked. Various
aspects of this interconnectivity\textbf{\ }are discussed by Rochet and
Tirole (1996), Freixas \textit{et al}. (2000), Pastor-Satorras and
Vespignani (2001), Leitner (2005), Egloff \textit{et al}. (2007), Allen and
Babus (2009), Wagner (2010), Haldane and May (2011), Steinbacher \textit{et
al}. (2014), Ladley (2013), Hurd (2015), among many others.

In the following subsection we specify dynamics for asset and liabilities,
which is consistent with a possibility of defaults by borrowers.\textbf{\ }

\subsection{Dynamics of Assets and Liabilities. Default Boundaries}

In the simplest possible case, the dynamics of assets and liabilities is
governed by the system of SDEs of the form%
\begin{equation}
\frac{dA_{i}\left( t\right) }{A_{i}\left( t\right) }=\mu dt+\sigma
_{i}dW_{i}\left( t\right) ,\ \ \ \frac{dL_{i}\left( t\right) }{L_{i}\left(
t\right) }=\mu dt,\ \ \ \frac{dL_{ij}\left( t\right) }{L_{ij}\left( t\right) 
}=\mu dt.  \label{Eq62}
\end{equation}%
where $\mu $ is growth rate, not necessarily risk neutral,\textbf{\ }$W_{i}$
are correlated Brownian motions, and $\sigma _{i}$ are corresponding
log-normal volatilities.

In a more general case, the corresponding dynamics can contain jumps, as
discussed in Lipton and Sepp (2009), or Itkin and Lipton (2015a, 2015b).%
\textbf{\ }Following Lipton and Sepp (2009)\textbf{, }we assume that
dynamics for firms' assets is given by

\begin{equation}
\frac{dA_{i}\left( t\right) }{A_{i}\left( t\right) }=\left( \mu -\kappa
_{i}\lambda _{i}\left( t\right) \right) dt+\sigma _{i}dW_{i}\left( t\right)
+\left( e^{J_{i}}-1\right) dN_{i}\left( t\right) ,  \label{Eq201}
\end{equation}%
where $N_{i}$ are Poisson processes independent of $W_{i}$, $\lambda _{i}$
are intensities of jump arrivals, $J_{i}$ are random jump amplitudes with
probability densities $\varpi _{i}\left( j\right) $, and $\kappa _{i}$ are
jump compensators, 
\begin{equation}
\kappa _{i}=\mathbb{E}\left\{ e^{J_{i}}-1\right\} .  \label{Eq202}
\end{equation}%
Since we are interested in studying consequences of default, it is enough to
assume that $J_{i}$ are negative exponential jumps, so that%
\begin{equation}
\varpi _{i}\left( j\right) =\left\{ 
\begin{array}{cc}
0, & j>0, \\ 
\vartheta _{i}e^{\vartheta _{i}j} & j\leq 0,%
\end{array}%
\right.  \label{Eq206}
\end{equation}%
with $\vartheta _{i}>0$. Diffusion processes $W_{i}$\textbf{\ }are
correlated in the usual way,%
\begin{equation}
dW_{i}\left( t\right) dW_{j}\left( t\right) =\rho _{ij}dt.  \label{Eq203}
\end{equation}%
Jump processes $N_{i}$\ are correlated in the spirit of Marshall-Olkin
(1967). We denote by $\Pi ^{\left( N\right) }$ the set of all subsets\textbf{%
\ }of $N$ names except for the empty subset $\left\{ \varnothing \right\} $,
and by $\pi $ a typical subset\textbf{.} With every $\pi $ we associate a
Poisson process $N_{\pi }\left( t\right) $ with intensity $\lambda _{\pi
}\left( t\right) $. Each $N_{i}\left( t\right) $ is projected on $N_{\pi
}\left( t\right) $ as follows%
\begin{equation}
N_{i}\left( t\right) =\sum_{\pi \in \Pi ^{\left( N\right) }}1_{\left\{ i\in
\pi \right\} }N_{\pi }\left( t\right) ,  \label{Eq204}
\end{equation}%
with%
\begin{equation}
\lambda _{i}\left( t\right) =\sum_{\pi \in \Pi ^{\left( N\right)
}}1_{\left\{ i\in \pi \right\} }\lambda _{\pi }\left( t\right) .
\label{Eq205}
\end{equation}%
Thus, for each bank we assume that there are both systemic and idiosyncratic
sources of jumps. In practice, it\textbf{\ }is sufficient to consider $N+1$
subsets of $\Pi ^{\left( N\right) }$, namely, the subset containing all
names, and subsets containing only one name at a time. For all other subsets
we put $\lambda _{\pi }=0$. If extra risk factors are needed, one can
include additional subsets representing particular industry sectors or
countries.

The simplest way of introducing default is to follow Merton's idea (Merton
1974) and to consider the process of final settlement at time $t=T$, see,
e.g., Webber and Willison (2011). However, given the highly regulated nature
of the banking business, it is hard to justify such a set-up. Accordingly,
we prefer to model the problem in the spirit of Black and Cox (1976) and
introduce continuous default boundaries\textbf{, }$\Lambda _{i}$\textbf{, }%
for $0\leq t\leq T$, which are defined as follows%
\begin{equation}
A_{i}\leq \Lambda _{i}=\left\{ 
\begin{array}{cc}
R_{i}\left( L_{i}+\hat{L}_{i}\right) -\hat{A}_{i}\equiv \Lambda _{i}^{<},\ \ 
& \ t<T, \\ 
L_{i}+\hat{L}_{i}-\hat{A}_{i}\equiv \Lambda _{i}^{=},\ \  & t=T,%
\end{array}%
\right.  \label{Eq63}
\end{equation}%
where\textbf{\ }$R_{i},0\leq R_{i}\leq 1$\textbf{\ }is the recovery rate. We
can think of $\Lambda _{i}$ as a function of external and mutual liabilities,%
\textbf{\ }$\mathbb{L=\{}L_{i},\hat{L}_{i}\}$, $\Lambda _{i}$=$\Psi
_{i}\left( \mathbb{L}\right) $.

If the $k$-th bank defaults at an intermediate time $t^{\prime }$, then the
capital of the remaining banks is depleted. We change indexation of the
surviving banks by applying the following function 
\begin{equation}
i\rightarrow i^{\prime }=\phi ^{k}\left( i\right) =\left\{ 
\begin{array}{ll}
i, & i<k, \\ 
i-1 & i>k.%
\end{array}%
\right.  \label{Eq64}
\end{equation}%
We also introduce the inverse function $\psi ^{k}$,%
\begin{equation}
i\rightarrow i^{\prime }=\psi ^{k}\left( i\right) =\left\{ 
\begin{array}{ll}
i, & i<k, \\ 
i+1 & i\geq k.%
\end{array}%
\right.  \label{Eq66}
\end{equation}%
The corresponding asset and liability matrices $\mathbb{A}^{\left( k\right)
} $, $\mathbb{L}^{\left( k\right) }$ assume the form%
\begin{equation}
\begin{array}{c}
\mathbb{A}^{\left( k\right) }=\left( A_{ij}^{\left( k\right) }\left(
t\right) \right) ,\ \ \ A_{ii}^{\left( k\right) }\left( t\right) =A_{\psi
^{k}\left( i\right) }\left( t\right) , \\ 
A_{ij}^{\left( k\right) }\left( t\right) =A_{\psi ^{k}\left( j\right) ,\psi
^{k}\left( i\right) }\left( t\right) , \\ 
\mathbb{L}^{\left( k\right) }=\left( L_{ij}^{\left( k\right) }\left(
t\right) \right) ,\ \ \ L_{ii}^{\left( k\right) }\left( t\right) =L_{\psi
^{k}\left( i\right) }\left( t\right) -L_{\psi ^{k}\left( i\right) ,k}\left(
t^{\prime }\right) +R_{k}L_{k,\psi ^{k}\left( i\right) }\left( t^{\prime
}\right) , \\ 
L_{ij}^{\left( k\right) }\left( t\right) =L_{\psi ^{k}\left( i\right) ,\psi
^{k}\left( i\right) }\left( t\right) , \\ 
t>t^{\prime }\text{,\ \ \ }i,j=1,...,N-1.%
\end{array}
\label{Eq65}
\end{equation}%
The corresponding default boundaries are given by%
\begin{equation}
A_{i}\leq \Lambda _{i}^{\left( k\right) }=\left\{ 
\begin{array}{cc}
R_{\psi ^{k}\left( i\right) }\left( L_{i}^{\left( k\right) }+\hat{L}%
_{i}^{\left( k\right) }-\hat{A}_{i}^{\left( k\right) }\right) , & t<T, \\ 
L_{i}^{\left( k\right) }+\hat{L}_{i}^{\left( k\right) }-\hat{A}_{i}^{\left(
k\right) },\ \ \  & t=T.%
\end{array}%
\right.  \label{Eq67}
\end{equation}%
$i\neq j$. It is clear that%
\begin{equation}
\Delta \Lambda _{i}^{\left( k\right) }=\Lambda _{i}^{\left( k\right)
}-\Lambda _{i}=\left\{ 
\begin{array}{cc}
\left( 1-R_{i}R_{k}\right) \hat{A}_{i}^{\left( k\right) }, & t<T, \\ 
\left( 1-R_{k}\right) \hat{A}_{i}^{\left( k\right) },\ \ \  & t=T.%
\end{array}%
\right.  \label{Eq68}
\end{equation}%
so that $\Delta \Lambda _{i}^{\left( k\right) }>0$ and the default
boundaries (naturally) move to the right.

\subsection{Terminal Settlement Conditions}

In order to formulate the terminal condition for the Kolmogorov equation, we
need to describe the settlement process at $t=T$ in the spirit of Eisenberg
and Noe (2001). Let $\vec{A}\left( T\right) $ be the vector of the terminal
external asset values. Since at time $T$ a full settlement is expected, we
assume that a particular bank will pay a fraction $\omega _{i}$ of its total
liabilities to its creditors (both external and inter-banks). If its assets
are sufficient to satisfy its obligations, then $\omega _{i}=1$, otherwise $%
0<\omega _{i}<1$. Thus, the settlement \ can be described by the following
system of equations%
\begin{equation}
\min \left( A_{i}\left( T\right) +\dsum\limits_{j}L_{ji}\omega _{j},L_{i}+%
\hat{L}_{i}\right) =\omega _{i}\left( L_{i}+\hat{L}_{i}\right) ,
\label{Eq69}
\end{equation}%
or equivalently%
\begin{equation}
\Phi _{i}\left( \vec{\omega}\right) \equiv \min \left( \frac{A_{i}\left(
T\right) +\dsum\limits_{j}L_{ji}\omega _{j}}{L_{i}+\hat{L}_{i}},1\right)
=\omega _{i}.  \label{Eq70}
\end{equation}%
It is clear that $\vec{\omega}$ is a fixed point of the mapping $\vec{\Phi}%
\left( \vec{\omega}\right) $,%
\begin{equation}
\vec{\Phi}\left( \vec{\omega}\right) =\vec{\omega}.  \label{Eq71}
\end{equation}%
Eisenberg and Noe have shown that $\vec{\Phi}\left( \vec{\omega}\right) $ is
a non-expanding mapping in the standard Euclidean metric, and formulated
conditions under which there is just one fixed point. We assume that these
conditions are satisfied, so that for each $\vec{A}\left( T\right) $ there
is a unique $\vec{\omega}\left( \vec{A}\left( T\right) \right) $ such that
the settlement is possible. There are no defaults provided that $\vec{\omega}%
=\vec{1}$, otherwise some banks default. Let $\vec{I}$ be a state\textbf{\ }%
indicator $\left( 0,1\right) $ vector of length $N$. Denote by $\mathcal{D}%
\left( \vec{I}\right) $ the following domain%
\begin{equation}
\mathcal{D}\left( \vec{I}\right) =\left\{ \left. \vec{A}\left( T\right)
\right\vert \omega _{i}\left( \vec{A}\left( T\right) \right) =\left\{ 
\begin{array}{cc}
1, & I_{i}=1 \\ 
<1 & I_{i}=0%
\end{array}%
\right. \right\} .  \label{Eq72}
\end{equation}%
In this domain the $i$-th bank survives provided that $I_{i}=1$, and
defaults otherwise. For example in the domain $\mathcal{D}\left(
1,...,1\right) $ all banks survive, while in the domain $\mathcal{D}\left(
0,1,...,1\right) $ the first bank defaults while all other survive, etc.

The actual terminal condition depends on the particular instrument under
consideration. If we are interested in the survival probability $Q$ of the
entire set of banks, we have%
\begin{equation}
Q\left( T,\vec{A}\right) =1_{\vec{A}\in \mathcal{D}\left( 1,...,1\right) }.
\label{Eq73}
\end{equation}%
For the marginal survival probability of the $i$-th bank we have%
\begin{equation}
Q\left( T,\vec{A}\right) =1_{\vec{A}\in \cup _{\vec{I}^{\left( i\right) }}%
\mathcal{D}\left( \vec{I}^{\left( i\right) }\right) },  \label{Eq74}
\end{equation}%
where $\vec{I}^{\left( i\right) }$ is the set of indicator vectors with $%
I_{i}=1$.

Thus far, we have introduced the stochastic dynamics for assets and
liabilities for a set of interconnected banks. These dynamics explicitly
allows for defaults of individual banks. Our framework reuses heavy
machinery originally developed in the context of credit derivatives. In
spite of being mathematical intense, such an approach is necessary to
quantitatively describe the financial sector as a manufacturer of credit.

\subsection{General Solution via Green's Function}

This Section is rather challenging mathematically and can easily be skipped
at first reading.

Our goal is to express general quantities of interest such as marginal
survival probabilities for individual banks and their joint survival
probability in terms of Green's function for the $N$-dimensional correlated
jump-diffusion process in a positive ortant.

As usual, it is more convenient to introduce normalized non-dimensional
variables. To this end, we define%
\begin{equation}
\bar{t}=\Sigma ^{2}t,\ \ \ \ X_{i}=\frac{\Sigma }{\sigma _{i}}\ln \left( 
\frac{A_{i}}{\Lambda _{i}^{<}}\right) ,\ \ \ \ \bar{\lambda}_{i}=\frac{%
\lambda _{i}}{\Sigma ^{2}},  \label{Eq75}
\end{equation}%
where 
\begin{equation}
\Sigma =\left( \dprod\limits_{i=1}^{N}\sigma _{i}\right) ^{1/N}.
\label{Eq76}
\end{equation}%
Thus,%
\begin{equation}
t=\frac{\bar{t}}{\Sigma ^{2}},\ \ \ \ A_{i}=\left( R_{i}\left( L_{i}+\hat{L}%
_{i}\right) -\hat{A}_{i}\right) e^{\sigma _{i}X_{i}/\Sigma }.  \label{Eq77}
\end{equation}%
The scaled default boundaries have the form%
\begin{equation}
X_{i}\leq M_{i}\left( t\right) =\left\{ 
\begin{array}{cc}
0\equiv M_{i}^{<},\ \  & t<T, \\ 
\frac{\Sigma }{\sigma _{i}}\ln \left( \frac{L_{i}\left( 0\right) +\hat{L}%
_{i}\left( 0\right) -\hat{A}_{i}\left( 0\right) }{R_{i}\left( L_{i}\left(
0\right) +\hat{L}_{i}\left( 0\right) \right) -\hat{A}_{i}\left( 0\right) }%
\right) \equiv M_{i}^{=},\ \  & t=T.%
\end{array}%
\right.  \label{Eq80}
\end{equation}%
The survival domain $\mathcal{D}\left( 1,...,1\right) $ is given by 
\begin{equation}
\mathcal{D}\left( 1,...,1\right) =\left\{ X_{i}>M_{i}^{=}\right\} ,
\label{Eq81}
\end{equation}%
Thus, we need to perform all our calculations in the positive cone $%
R_{+}^{\left( N\right) }$.

The dynamics of $\vec{X}=\left( X_{1},...,X_{N}\right) $ is governed by the
following equations%
\begin{eqnarray}
dX_{i} &=&-\left( \frac{\sigma _{i}}{2\Sigma }+\kappa _{i}\ \bar{\lambda}%
_{i}\right) d\bar{t}+dW_{i}\left( \bar{t}\right) +\frac{\Sigma }{\sigma _{i}}%
J_{i}dN_{i}\left( \bar{t}\right)  \label{Eq77a} \\
&\equiv &\xi _{i}d\bar{t}+dW_{i}\left( \bar{t}\right) +\zeta
_{i}J_{i}dN_{i}\left( \bar{t}\right) .  \notag
\end{eqnarray}%
Below we omit bars for the sake of brevity and rewrite Eq. (\ref{Eq77a}) in
the form:%
\begin{equation}
dX_{i}=\xi _{i}dt+dW_{i}\left( t\right) +\zeta _{i}J_{i}dN_{i}\left(
t\right) ,  \label{Eq78}
\end{equation}%
The corresponding Kolmogorov backward operator has the form 
\begin{eqnarray}
\mathcal{L}^{\left( N\right) }f &=&\dsum\limits_{i=1}^{N}\left( \frac{1}{2}%
\dsum\limits_{j=1}^{N}\rho _{ij}f_{X_{i}X_{j}}+\xi _{i}f_{X_{i}}\right)
\label{Eq79} \\
&&+\tsum\limits_{\pi \in \Pi ^{\left( N\right) }}\lambda _{\pi
}\prod\limits_{i\in \pi }\mathcal{J}_{i}f\left( \vec{X}\right)
-\tsum\limits_{\pi \in \Pi ^{\left( N\right) }}\lambda _{\pi }f\left( \vec{X}%
\right)  \notag \\
&\equiv &\frac{1}{2}\Delta _{\rho }f+\xi \cdot \nabla f+\mathcal{J}%
f-\upsilon f,  \notag
\end{eqnarray}%
where%
\begin{equation}
\mathcal{J}\left( \vec{X}\right) \mathcal{=}\tsum\limits_{\pi \in \Pi
^{\left( N\right) }}\lambda _{\pi }\prod\limits_{i\in \pi }\mathcal{J}%
_{i}f\left( \vec{X}\right) ,  \label{Eq220}
\end{equation}%
\begin{equation}
\mathcal{J}_{i}f\left( \vec{X}\right) =\varsigma _{i}\int_{0}^{X_{i}}f\left(
X_{1},...,X_{i}-j,...X_{N}\right) e^{-\varsigma _{i}j}dj,  \label{Eq207}
\end{equation}%
and $\varsigma _{i}=\sigma _{i}\vartheta _{i}/\Sigma $.

We can formulate a typical pricing equation in the positive cone $%
R_{+}^{\left( N\right) }$. We have%
\begin{equation}
\partial _{t}V\left( t,\vec{X}\right) +\mathcal{L}^{\left( N\right) }V\left(
t,\vec{X}\right) =\chi \left( t,\vec{X}\right) ,  \label{Eq208}
\end{equation}%
\begin{equation}
V\left( t,\vec{X}_{0,k}\right) =\phi _{0,k}\left( t,\vec{Y}\right) ,\ \ \ \
\ V\left( t,\vec{X}_{\infty ,k}\right) =\phi _{\infty ,k}\left( t,\vec{Y}%
\right) ,  \label{Eq209}
\end{equation}%
\begin{equation}
V\left( T,\vec{X}\right) =\psi \left( \vec{X}\right) ,  \label{Eq210}
\end{equation}%
where $\vec{X}$, $\vec{X}_{0,k}$, $\vec{X}_{\infty ,k}$, $\vec{Y}_{k}$ are $%
N $ and $N-1$ dimensional vectors, respectively,%
\begin{equation}
\begin{array}{c}
\vec{X}=\left( x_{1},...,x_{k},...x_{N}\right) , \\ 
\vec{X}_{0,k}=\left( x_{1},...,\underset{k}{0},...x_{N}\right) , \\ 
\vec{X}_{\infty ,k}=\left( x_{1},...,\underset{k}{\infty },...x_{N}\right) ,
\\ 
\vec{Y}_{k}=\left( x_{1},...x_{k-1},x_{k+1},...x_{N}\right) .%
\end{array}
\label{Eq211}
\end{equation}%
Here $\chi \left( t,\vec{X}\right) $, $\phi _{0,k}\left( t,\vec{y}\right) $, 
$\phi _{\infty ,k}\left( t,\vec{y}\right) $, $\psi \left( \vec{X}\right) $
are known functions, which are contract specific. For instance, for the
joint survival probability $Q\left( t,\vec{X}\right) $ we have%
\begin{equation}
\chi \left( t,\vec{X}\right) =0,\ \ \ \phi _{0,k}\left( t,\vec{Y}\right)
=\phi _{\infty ,k}\left( t,\vec{Y}\right) =0,\ \ \psi \left( \vec{X}\right)
=1_{\vec{X}\in \mathcal{D}\left( 1,...,1\right) }.\   \label{Eq212}
\end{equation}

The corresponding adjoint operator is%
\begin{equation}
\mathcal{L}^{\left( N\right) \dagger }g\left( \vec{X}\right) =\frac{1}{2}%
\Delta _{\rho }g-\xi \cdot \nabla g+\mathcal{J}^{\dagger }g-\upsilon g,
\label{Eq213}
\end{equation}%
where%
\begin{equation}
\mathcal{J}^{\dagger }g\left( \vec{X}\right) =\tsum\limits_{\pi \in \Pi
^{\left( N\right) }}\lambda _{\pi }\prod\limits_{i\in \pi }\mathcal{J}%
_{i}^{\dagger }g\left( \vec{X}\right) ,  \label{Eq214}
\end{equation}%
\begin{equation}
\mathcal{J}_{i}^{\dagger }g\left( \vec{X}\right) =\varsigma
_{i}\int_{0}^{\infty }g\left( X_{1},...,X_{i}+j,...x_{N}\right)
e^{-\varsigma _{i}j}dj,  \label{Eq215}
\end{equation}%
It is easy to check that%
\begin{equation}
\int_{R_{+}^{\left( N\right) }}\left[ \mathcal{J}_{i}f\left( \vec{X}\right)
g\left( \vec{X}\right) -f\left( \vec{X}\right) \mathcal{J}_{i}^{\dagger
}g\left( \vec{X}\right) \right] d\vec{X}=0.  \label{Eq216}
\end{equation}

We solve Eqs (\ref{Eq208})-(\ref{Eq210}) by introducing the Green's function 
$G\left( t,\vec{X}\right) $, or, more explicitly, $G\left( t,\vec{x};0,\vec{X%
}^{\prime }\right) $, such that%
\begin{equation}
\partial _{t}G\left( t,\vec{x}\right) -\mathcal{L}^{\left( N\right) \dagger
}G\left( t,\vec{x}\right) =0,  \label{Eq217}
\end{equation}%
\begin{equation}
G\left( t,\hat{X}_{0}^{\left( k\right) }\right) =0,\ \ \ \ G\left( t,\hat{X}%
_{\infty }^{\left( k\right) }\right) =0,  \label{Eq218}
\end{equation}%
\begin{equation}
G\left( 0,\vec{X}\right) =\delta \left( \vec{X}-\vec{X}^{\prime }\right) .
\label{Eq219}
\end{equation}%
It is clear that%
\begin{equation}
\left( VG\right) _{t}+\mathcal{L}VG-V\mathcal{L}^{\dag }G=\chi G.
\label{Eq85}
\end{equation}%
Some relatively simple algebra yields%
\begin{equation}
\left( VG\right) _{t}+\nabla \cdot \left( \vec{F}\left( V,G\right) \right) +%
\mathcal{J}VG-V\mathcal{J}^{\dagger }G=\chi G,  \label{Eq86}
\end{equation}%
where%
\begin{eqnarray}
\vec{F} &\mathbf{=}&\left( F_{1},...,F_{i},...F_{N}\right)  \label{Eq87} \\
&=&\left( F_{1}^{\left( 1\right) },...,F_{i}^{\left( 1\right)
},...F_{N}^{\left( 1\right) }\right) +\left( F_{1}^{\left( 2\right)
},...,F_{i}^{\left( 2\right) },...F_{N}^{\left( 2\right) }\right)  \notag \\
&\equiv &\vec{F}^{\left( 1\right) }+\vec{F}^{\left( 2\right) },  \notag \\
F_{i}^{\left( 1\right) } &=&\frac{1}{2}V_{X_{i}}G+\xi _{i}VG+\left(
\dsum\limits_{j<i}\rho _{ij}V_{X_{j}}\right) G,  \notag \\
F_{i}^{\left( 2\right) } &=&-\frac{1}{2}VG_{X_{i}}-V\left(
\dsum\limits_{j>i}\rho _{ij}G_{X_{j}}\right) .  \notag
\end{eqnarray}%
Green's theorem yields 
\begin{eqnarray}
V\left( 0,\vec{X}^{\prime }\right) &=&\int_{R_{+}^{\left( N\right) }}\psi
\left( \vec{X}\right) G\left( T,\vec{X}\right) d\vec{X}  \label{Eq88} \\
&&+\dsum\limits_{k}\int_{0}^{T}dt\int_{R_{+}^{\left( N-1\right) }}\phi
_{0,k}\left( t,\vec{Y}\right) g_{_{K}}\left( t,\vec{Y}\right) d\vec{Y} 
\notag \\
&&-\int_{0}^{T}\int_{R_{+}^{\left( N\right) }}\chi \left( t,\vec{X}\right)
G\left( t,\vec{X}\right) dtd\vec{X},  \notag
\end{eqnarray}%
where%
\begin{equation}
g_{k}\left( t,\vec{Y}\right) =\frac{1}{2}G_{X_{k}}\left( t,X_{1},...,%
\underset{k}{0},...,X_{N}\right) .  \label{Eq221}
\end{equation}%
Thus, in order to solve the backward pricing problem with nonhomogeneous
right hand side and boundary conditions, it is sufficient solve the forward
propagation problem for Green's function with homogeneous right hand side
and boundary conditions.

In particular, for the joint survival probability, we have%
\begin{equation}
Q\left( 0,\vec{X}^{\prime }\right) =\underset{\vec{X}\in \mathcal{D}\left(
1,...,1\right) }{\dint }G\left( T,\vec{X}\right) dX_{1}...dX_{N}.
\label{Eq89}
\end{equation}%
Similarly, for the marginal survival probability of the first bank, say, we
have%
\begin{eqnarray}
Q_{1}\left( 0,\vec{X}^{\prime }\right) &=&\underset{\vec{X}\in \mathcal{D}%
\left( 0,...,1\right) }{\dint }G\left( T,\vec{X}\right) d\vec{X}
\label{Eq90a} \\
&&+\dsum\limits_{k}\int_{0}^{T}dt\int_{R_{+}^{\left( N-1\right)
}}Q_{1}\left( t,\vec{Y}\right) g_{_{K}}\left( t,\vec{Y}\right) d\vec{Y}. 
\notag
\end{eqnarray}

\section{Banks' Balance Sheet Optimization\label{BSOptimization}}

This Section is aimed at increasing the granularity of our model. Let's
recall that first we considered a simple economy as a whole and assumed that
it is driven by stochastic demand for goods and money, and described the
corresponding monetary circuit. In this framework, physical goods and money
are treated in a uniform fashion. Next, we moved on to a more granular level
and described a system of interlinked banks that create money by
accommodating external changes in demand for it. Now, we have reached the
most granular level of our theory, and consider an individual bank. We
emphasize that MMC theory described in this paper is a top-down theory.
However, once major consistent patterns from the overall economy are traced
to the level of an individual bank, the consequences for the bank
profitability and risk management are hard to overestimate.

Numerous papers and monographs deal with various aspects of the bank balance
sheet optimization problem. Here we mention just a few. Kusy and Ziemba
(1986) develop a multi-period stochastic linear programming model for
solving a small bank asset and liability management (ALM) problem. dos Reis
and Martins (2001) develop an optimization model and use it to choose the
optimal categories of assets and liabilities to form a balance sheet of a
profitable and sound bank. In a series of papers, Petersen and coauthors
analyze bank management via stochastic optimal control and suggest an
optimal portfolio choice and rate of bank capital inflow that keep the loan
level close to an actuarially determined reference process, see, e.g.,
Mukuddem-Petersen and Petersen (2006). Dempster \textit{et al.} (2009) show
how to use dynamic stochastic programming in order to perform optimal
dynamic ALM over long time horizons; their ideas can be expanded to cover a
bank balance sheet optimization. Birge and Judice (2013) propose a dynamic
model which encompasses the main risks in balance sheets of banks and use it
to simulate bank balance sheets over time given their lending strategy and
to determine optimal bank ALM strategy endogenously. Halaj (2012) proposes a
model of optimal structure of bank balance sheets incorporating strategic
and optimizing behavior of banks under stress scenarios. Astic and Tourin
(2013) propose a structural model of a financial institution investing in
both liquid and illiquid assets and use stochastic control techniques to
derive the variational inequalities satisfied by the value function and
compute the optimal allocations of assets. Selyutin and Rudenko (2013)
develop a novel approach to ALM problem based on the transport equation for
loan and deposit dynamics.

To complement the existing literature, we develop {a framework for
optimizing enterprise business portfolio by mathematically analyzing
financial and risk metrics\ across various economic scenarios, with an
overall objective to maximize risk adjusted return, while staying within
various constraints. Regulations impose multiple capital requirements and
constraints\ on the banking industry (such as B3S and B3A capital ratios,
Leverage Ratios, Liquidity Coverage Ratios, etc.).}

The economic objective of the balance sheet optimization for an individual
bank is to choose the level of Loans, Deposits, Investments, \ Debt and
Capital in such a way as to satisfy Basel III rules and, at the same time,
maximize cash flows attributable to shareholders. Balance sheet optimization
boils down to solving a very involved Hamilton Jacobi Bellman problem.%
\textbf{\ }The optimization problem {can be formulated in two ways: (a) }%
Optimize{\textbf{\ }cashflows without using a\textbf{\ }}risk preference {%
utility function, or, equivalently, \textbf{\ }being indifferent to the
probability of loss vs. profits; (b) }Introduce a {utility function }into
the optimization problem and solve it in the spirit of Merton's optimal
consumption problem.{\ }Although, as a rule, balance sheet optimization has
to be done numerically, occasionally, depending on the chosen utility
function, a semi-analytical solution can be obtained.

\subsection{Notations and Main Variables}

Let us introduce key notation. By necessity, we have to reuse some of the
symbols used earlier; we hope this will not confuse the reader. Bank's
assets in increasing order of liquidity have the form

\begin{eqnarray*}
&&X_{k}^{\pi }\text{, outstanding loans with maturity }T_{k}\text{ and
quality }p, \\
&&I\text{, investments in stocks and bonds,} \\
&&C\text{, cash.}
\end{eqnarray*}%
We {assume that $T_{1}<...<T_{k}<...<T_{K}$, and $p=1,...,P$. Quality of
loans is determined by various factors, such as the rating of the borrower,
collateralization, etc.}

Bank's liabilities in increasing order of stickiness have the form

\begin{eqnarray*}
&&D\text{, deposits,} \\
&&Y_{l}^{q}\text{, outstanding debts with maturity }T_{l}\text{ and quality }%
q, \\
&&E\text{, equity (or capital).}
\end{eqnarray*}%
We {assume that $T_{1}<...<T_{l}<...<T_{L}$, and $q=1,...,Q$. Quality of
borrowings is determined by various factors, such as its seniority,
collateralization, etc.}

Assets and liabilities have the following properties: {(a) Loans and debts
are characterized by their repayment/loss rates }$\lambda _{k}^{p}${\ and }$%
\mu _{l}^{q}${, and interest rates }$\nu _{k}^{p}${\ and }$\xi _{l}^{q}$;
(b) {Similarly, for deposits we have rates }$\alpha $ and $\beta ,${\
respectively; (c) Finally, for investments the corresponding}\textbf{\ }{%
growth rates are stochastic and have the form }$r-\zeta +\sigma \chi \left(
t\right) $,{\ where $r$ is the expected growth rate, $\zeta $ is the
dividend rate, $\sigma $ is the volatility of returns}\ on investments{, and 
}$\chi \left( t\right) =dW\left( t\right) /dt${\ is white noise, or
"derivative" of the standard Brownian motion, so that%
\begin{equation}
dI=\left( r-\zeta \right) Idt+\sigma IdW.  \label{Eq91}
\end{equation}%
\qquad \qquad }

Balance Sheet Balancing Equation {has the form:%
\begin{equation}
\sum_{k,p}X_{k}^{p}+I+C=D+\sum_{l,q}Y_{l}^{q}+E.  \label{Eq92}
\end{equation}%
Below we omit sub- and superscripts for brevity and rewrite the equation of
balance as follows:%
\begin{equation}
X+I+C-D-Y-E=0.  \label{Eq93}
\end{equation}%
}

There are several controls and levers for determining general direction of
the bank: {(a) rates $\phi \left( t\right) $ at which new loans are issued;
(b) rates $\psi \left( t\right) $ at which new borrowings are obtained; (c)
rate $\omega \left( t\right) $ at which new investments are made; (d) rate $%
\pi \left( t\right) $ at which new deposits are acquired; (e) rate $\delta
\left( t\right) $ at which money is returned to shareholders in the form of
dividends or }share {buy-backs. If }${\delta \left( t\right) <0}${, then new
stock is issued. Of course, dividends should not be paid when new shares are
issued.}

The evolution of the bank's assets and liabilities is governed by the
following equations:

\begin{eqnarray}
X^{\prime }\left( t\right) &=&-\lambda X\left( t\right) +\Phi \left(
t\right) ,  \label{Eq94} \\
I^{\prime }\left( t\right) &=&\left( r-\zeta +\sigma \chi \left( t\right)
\right) I\left( t\right) +\omega \left( t\right) ,  \notag \\
C^{\prime }\left( t\right) &=&-X^{\prime }\left( t\right) +\nu X\left(
t\right) +\zeta I\left( t\right) -\omega \left( t\right)  \notag \\
&&+D^{\prime }\left( t\right) -\beta D\left( t\right) +Y^{\prime }\left(
t\right) -\xi Y\left( t\right) -\delta \left( t\right)  \notag \\
&=&\left( \lambda +\nu \right) X\left( t\right) -\Phi \left( t\right) +\zeta
I\left( t\right) -\omega \left( t\right)  \notag \\
&&-\left( \alpha +\beta \right) D\left( t\right) +\pi \left( t\right)
-\left( \mu +\xi \right) Y\left( t\right) +\Psi \left( t\right) -\delta
\left( t\right) ,  \notag
\end{eqnarray}%
and%
\begin{eqnarray}
D^{\prime }\left( t\right) &=&-\alpha D\left( t\right) +\pi \left( t\right) ,
\label{Eq94a} \\
Y^{\prime }\left( t\right) &=&-\mu Y\left( t\right) +\Psi \left( t\right) , 
\notag \\
E^{\prime }\left( t\right) &=&\nu X\left( t\right) +I^{\prime }\left(
t\right) +\zeta I\left( t\right) -\omega \left( t\right) -\beta D\left(
t\right) -\xi Y\left( t\right) -\delta \left( t\right)  \notag \\
&=&\nu X\left( t\right) +\left( r+\sigma \chi \left( t\right) \right)
I\left( t\right) -\beta D\left( t\right) -\xi Y\left( t\right) -\delta
\left( t\right) ,  \notag
\end{eqnarray}%
respectively. {Here, }for convenience, instead of $\phi \left( t\right) $\
and $\psi \left( t\right) $\ we use ${\Phi \left( t\right) }$ and ${\Psi
\left( t\right) }$, defined as follows {\ }%
\begin{equation}
\begin{array}{c}
{\Phi \left( t\right) }{=\phi \left( t\right) -e^{-\lambda T}\phi \left(
t-T\right) ,} \\ 
{\Psi \left( t\right) }{=\psi \left( t\right) -e^{-\mu T}\psi \left(
t-T\right) ,}%
\end{array}
\label{Eq95}
\end{equation}%
respectively.

On the bank's asset side, outstanding loans decay deterministically
proportionally to their repayment rate and increase due to new loans issued
less amortized old loans repaid. Existing investments grow stochastically as
in Eq. (\ref{Eq91}) and are complemented by new investments. Changes in cash
balances are influenced by several factors. On the one hand, prepaid loans,
interest charged on outstanding loans, dividends on investments, new
deposits, and new borrowings positively contribute to cash balances. On the
other hand, new investments, interest paid on deposits and borrowings,
withdrawn deposits and losses on lending, as well as money returned to the
shareholders as dividends and/or share buy-backs lead to reduction in the
bank cash position.

On the bank's liability side, deposits decay deterministically
proportionally to their withdrawal rate and increase due to new deposits
coming in.\textbf{\ }Outstanding bank's debts decay deterministically at
their repayment rate, and increase due to new borrowings less amortized old
debts repaid.\textbf{\ }Similarly to changes in cash on the asset side,
changes in capital (equity) on the liability side are positively affected by
the interest paid on outstanding loans, stochastic returns on investments
(including dividends), and negatively affected by interest paid on deposits,
borrowings, and dividends paid to the shareholders.

{Balancing equation (\ref{Eq93}) after differentiation} becomes{%
\begin{equation}
X^{\prime }+I^{\prime }+C^{\prime }-D^{\prime }-Y^{\prime }-E^{\prime }=0.
\label{Eq96}
\end{equation}%
}and is identically satisfied by virtue of Eqs (\ref{Eq94}), (\ref{Eq94a})
since{%
\begin{equation}
\begin{array}{c}
X^{\prime }+I^{\prime }+C^{\prime }-D^{\prime }-Y^{\prime }-E^{\prime }= \\ 
X^{\prime }+I^{\prime }-X^{\prime }+\nu X+\zeta I-\omega +D^{\prime }-\beta
D+Y^{\prime }-\xi Y-\delta \\ 
-D^{\prime }-Y^{\prime }-\nu X-I^{\prime }-\zeta I+\omega +\beta D+\xi
Y+\delta =0.%
\end{array}
\label{Eq96a}
\end{equation}%
}

\subsection{Optimization Problem}

The cashflow $CF\left( T\right) $ attributable to the common equity up to
and including some terminal time $T$ is determined by the discounted
expected value of change in equity plus the discounted value of money
returned to shareholders over a given time period. By using Eqs (\ref{Eq94a}%
), $CF\left( T\right) $\ can be calculated as follows:{%
\begin{equation}
\begin{array}{c}
CF\left( T\right) =e^{-RT}\mathbb{E}\left\{ E\left( T\right) \right\}
-E\left( 0\right) +\int_{0}^{T}e^{-Rt}\delta \left( t\right) dt \\ 
\ \ \ \ \ =e^{-RT}\mathbb{E}\left\{ \int_{0}^{T}E^{\prime }\left( t\right)
+e^{-R\left( t-T\right) }\delta \left( t\right) dt\right\} \\ 
\ \ \ \ \ \ \ \ \ =e^{-RT}\mathbb{E}\left\{ \int_{0}^{T}\left( \nu X\left(
t\right) +\left( r+\sigma \chi \left( t\right) \right) I\left( t\right)
\right. \right. \\ 
\ \ \ \ \ \ \ \ \ \ \ \ \ \ \ \ \ \ \left. \left. -\beta D\left( t\right)
-\xi Y\left( t\right) -\delta \left( t\right) +e^{-R\left( t-T\right)
}\delta \left( t\right) \right) dt\right\} \\ 
\ \ \ \ \ \ \ \ \ \ \ \ =e^{-RT}\int_{0}^{T}\left( \nu X\left( t\right)
+rJ\left( t\right) -\beta D\left( t\right) -\xi Y\left( t\right) \right. \\ 
\left. +\left( e^{-R\left( t-T\right) }-1\right) \delta \left( t\right)
\right) dt.%
\end{array}
\label{Eq97}
\end{equation}%
Here $R$ is the discount rate, and $J\left( t\right) $ is the expected value
of }investments {$I\left( t\right) $ with dividends reinvested. The }%
deterministic {governing equation for }$J$ {has the form:%
\begin{equation}
J^{\prime }\left( t\right) =rJ\left( t\right) +\omega \left( t\right)
\label{Eq98}
\end{equation}%
}Accordingly, in order to optimize the balance sheet at the most basic
level, we need to maximize $CF\left( T\right) $, viewed as a functional
depending on $\phi \left( t\right) ,\omega \left( t\right) ,\pi \left(
t\right) ,\psi \left( t\right) ,$ and $\delta \left( t\right) $:{%
\begin{equation}
CF\left( T\right) \underset{\phi \left( t\right) ,\omega \left( t\right)
,\pi \left( t\right) ,\psi \left( t\right) ,\delta \left( t\right) }{%
\rightarrow }\max .  \label{Eq99}
\end{equation}%
}However, this optimization problem\ is subject to various regulatory\
constraints, such as capital, liquidity, leverage, etc., some of which are
explicitly described below. Clearly, the problem has numerous degrees of
freedom, which can be reduced somewhat by assuming, for example, that $\phi
\left( t\right) ,\omega \left( t\right) ,\pi \left( t\right) ,\psi \left(
t\right) ,\delta \left( t\right) $ are time independent.

\subsection{Capital Constraints}

{Regulatory capital calculations are fairly complicated. }They are based on
systematizing and aggregating bank portfolio's assets into risk groups and
assigning risk weights to each group. Therefore, for determining{\ Risk
Weighted Assets (RWAs), it is necessary to classify loans and investments as
Held To Maturity (HTM), Available For Sale (AFS), or belonging to the
Trading Book (TB).}

{We start with HTM and AFS bonds. We can use either the standard model (SM),
or an internal rating based model (IRBM)\textbf{. }SM represents RWA in the
form:%
\begin{equation}
RWA_{SM}=rwa_{SM}\cdot X,  \label{Eq109}
\end{equation}%
where the weights }$rwa_{SM}=\left( r{wa_{SM,k}^{p}}\right) ${\ are
regulatory prescribed, and}%
\begin{equation}
rwa_{SM}\cdot X=\dsum\limits_{k,p}rwa_{SM,k}^{p}X_{k}^{p}.  \label{Eq109a}
\end{equation}%
{\ Alternatively, IRBM\textbf{\ }}provides{\ the following expression for
the RWAs:%
\begin{equation}
RWA_{IRBM}=rwa_{IRBM}\cdot X,  \label{Eq110}
\end{equation}%
where the weights }$rwa_{IRBM}${$=$}$\left( r{wa_{IRBM,k}^{p}}\right) ${\
are given by relatively complex formulas, which are omitted for brevity. In
both cases, the corresponding regulatory capital is given by }%
\begin{equation}
{K^{\left( 1\right) }=\kappa RWA.}  \label{Eq111}
\end{equation}%
Additional amounts of capital $K^{\left( 2\right) },K^{\left( 3\right)
},K^{\left( 4\right) }$ are required to cover counterparty, operational and
market risks, respectively, so that the total amount of capital the bank
needs to hold\textbf{\ }is given by%
\begin{equation}
K=K^{\left( 1\right) }+K^{\left( 2\right) }+K^{\left( 3\right) }+K^{\left(
4\right) }.  \label{Eq112}
\end{equation}%
{It is clear that for a bank to be a going concern, the following inequality
has to be satisfied%
\begin{equation}
E-K>0.  \label{Eq113}
\end{equation}%
}

\subsection{Liquidity Constraints}

We formulate liquidity constraints in terms of the following quantities:

(a) Required Stable Funding (RSF)

\begin{equation}
RSF=rsf_{X}\cdot X+rsf_{I}\cdot I+0\cdot C;  \label{Eq100}
\end{equation}

(b) {Available Stable Funding (ASF)}

{%
\begin{equation}
ASF=asf_{D}\cdot D+asf_{Y}\cdot Y+1\cdot E.  \label{Eq101}
\end{equation}%
}Here $rsf_{X}=\left( rsf_{k}^{p}\right) $, and 
\begin{equation}
rsf_{X}\cdot X=\dsum\limits_{k,p}rsf_{k}^{p}X_{k}^{p}.  \label{Eq102}
\end{equation}

{In addition, we define: }

(c) {Stylized 30 day cash outflows (CO)}

\begin{equation}
CO=co_{D}\cdot D+co_{Y}\cdot Y+0\cdot E;  \label{Eq103}
\end{equation}

(d) {Stylized 30 day cash inflows (CI):}

{%
\begin{equation}
CI=ci_{X}\cdot X+ci_{I}\cdot I+1\cdot C.  \label{Eq104}
\end{equation}%
}Here the weights $rsf_{X}$, $rsf_{I}$,$asf_{D}$, $asf_{Y}$, $co_{D}$, $%
co_{Y}$, $ci_{X}$, $ci_{I}$ are prescribed by the regulators.

In order to comply with Basel III requirements, it is necessary to have:%
\begin{equation}
ASF>RSF,  \label{Eq105}
\end{equation}%
{%
\begin{equation}
CI>CO,  \label{Eq106}
\end{equation}%
}or equivalently,%
\begin{equation}
-rsf\cdot X-rsf_{I}\cdot I+asf_{D}\cdot D+asf\cdot Y+E>0,  \label{Eq107}
\end{equation}%
{%
\begin{equation}
ci\cdot X+ci_{I}\cdot I+C-co_{D}\cdot D-co\cdot Y>0.  \label{Eq108}
\end{equation}%
}In words, Eqs (\ref{Eq107}) and (\ref{Eq108}) indicate that having large
amounts of equity, $E$\ and capital, $C$\ is beneficial for the bank's
liquidity position (but not for its earnings!).

\subsection{Mathematical Formulation: General Optimization Problem\label%
{GeneralOptimization}}

A {general optimization problem can be formulated in terms of independent
variables $X,I,C,D,Y$ defined in the multi-dimensional domain given by the
corresponding constraints }

{There are adjacent domains where complementary variational inequalities are
satisfied. The corresponding HJB equation reads: 
\begin{equation}
\underset{\phi ,\omega ,\pi ,\psi ,\delta }{\max }\left\{ 
\begin{array}{c}
V_{t}+\frac{1}{2}\sigma ^{2}I^{2}V_{II}+\left( -\lambda X+\Phi \right)
V_{X}+\left( r-\zeta \right) IV_{I} \\ 
+\left( \left( \lambda +\nu \right) X-\Phi +\zeta I-\omega \right. \\ 
\left. -\left( \alpha +\beta \right) D+\pi -\left( \mu +\xi \right) Y+\Psi
\right) V_{C} \\ 
+\left( -\alpha D+\pi \right) V_{D}+\left( -\mu Y+\Psi \right) V_{Y}-RY, \\ 
1-V_{C}%
\end{array}%
\right\} =0.  \label{Eq114}
\end{equation}%
}

{In the limit of $T\rightarrow \infty $ the problem simplifies to (but still
remains very complex):%
\begin{equation}
\underset{\phi ,\omega ,\pi ,\psi ,\delta }{\max }\left\{ 
\begin{array}{c}
\frac{1}{2}\sigma ^{2}I^{2}V_{II}+\left( -\lambda X+\Phi \right)
V_{X}+\left( r-\zeta \right) IV_{I} \\ 
+\left( \left( \lambda +\nu \right) X-\Phi +\zeta I-\omega \right. \\ 
\left. -\left( \alpha +\beta \right) D+\pi -\left( \mu +\xi \right) Y+\Psi
\right) V_{C} \\ 
+\left( -\alpha D+\pi \right) V_{D}+\left( -\mu Y+\Psi \right) V_{Y}-RY, \\ 
1-V_{C}%
\end{array}%
\right\} =0.  \label{Eq115}
\end{equation}%
}

\subsection{Mathematical Formulation: Simplified Optimization Problem\label%
{SimplifiedOptimization}}

Instead of dealing with several independent variables, $X,...,Y$, we
concentrate on the equity portion of the capital structure, $E$,\ which
follows the effective evolution equation:

\begin{equation}
dE=\left( \mu -d\right) dt+\sigma dW-J_{1}dN_{1}-J_{2}dN_{2},  \label{Eq116}
\end{equation}%
where $\mu $ is the accumulation rate, $d$ is the dividend rate, which we
wish to optimize, $\sigma $ is the volatility of earnings, $W$ is Brownian
motion, $N_{1,2}$ are two independent Poisson processes with frequencies $%
\lambda _{1,2}$, and $J_{1,2}$ are exponentially distributed jumps, $%
J_{i}\sim \delta _{i}\exp \left( -\delta _{i}j\right) $. The choice of the
jump-diffusion dynamics with two independent Poisson drivers reflects the
fact that the growth of the bank's equity is determined by retained profits,
which are governed by an arithmetic Browinian motion, and negatively
affected by two types of jumps, namely, more frequent (but slightly less
dangerous due to potential actions of the central bank) liquidity jumps
represented by $N_{2}$, and less frequent (but much more dangerous) solvency
jumps represented by $N_{1}$. Accordingly, $\lambda _{1}>\lambda _{2}$, and $%
\delta _{1}<\delta _{2}$. Below we assume that the dividend rate is
potentially unlimited, so that a lump sum can be paid instantaneously. A
similar problem with just one source of jumps has been considered in the
context of an insurance company interested in maximization of its dividend
pay-outs (see, e.g., Taksar 2000 and Belhaj 2010 and references therein).

The bank defaults when $E$ crosses zero. We shall see shortly that it is
optimal for the bank not to pay any dividend until $E$ reaches a certain
optimal level $E^{\ast }$, and when this level is reached, to pay all the
excess equity in dividends at once. With all the specifics in mind, the
dividend optimization problem (\ref{Eq114}) can be mathematically formulated
as follows\textbf{\ }{%
\begin{equation}
\underset{d}{\max }\left\{ 
\begin{array}{c}
V_{t}+\frac{1}{2}\sigma ^{2}V_{EE}+\left( \mu -d\right) V_{E}-\left(
R+\lambda _{1}+\lambda _{2}\right) V \\ 
+\lambda _{1}\delta _{1}\int_{0}^{E}V\left( E-J_{1}\right) e^{-\delta
_{1}J_{1}}dJ_{1} \\ 
+\lambda _{2}\delta _{2}\int_{0}^{E}V\left( E-J_{2}\right) e^{-\delta
_{2}J_{2}}dJ_{2}+d%
\end{array}%
\right\} =0,  \label{Eq117}
\end{equation}%
}%
\begin{equation}
V\left( T,E\right) =E,\ \ \ E\geq 0,  \label{Eq118}
\end{equation}%
\begin{equation}
V\left( t,0\right) =0,\ \ \ \ 0\leq t\leq T.  \label{Eq119}
\end{equation}%
Solving Eq. (\ref{Eq117}) supplemented with terminal and boundary conditions
(\ref{Eq118})-(\ref{Eq119}) is equivalent to solving the following
variational inequality:{%
\begin{equation}
\max \left\{ 
\begin{array}{c}
V_{t}+\frac{1}{2}\sigma ^{2}V_{EE}+\mu V_{E}-\left( R+\lambda _{1}+\lambda
_{2}\right) V \\ 
+\lambda _{1}\delta _{1}\int_{0}^{E}V\left( E-J_{1}\right) e^{-\delta
_{1}J_{1}}dJ_{1} \\ 
+\lambda _{2}\delta _{2}\int_{0}^{E}V\left( E-J_{2}\right) e^{-\delta
_{2}J_{2}}dJ_{2}, \\ 
1-V_{E}%
\end{array}%
\right\} =0,  \label{Eq120}
\end{equation}%
augmented with conditions\ (\ref{Eq118}), (\ref{Eq119}). }We use generic
notation to rewrite Eq. (\ref{Eq120}) as follows:{%
\begin{equation}
\max \left\{ V_{t}+a_{2}V_{EE}+a_{1}V_{E}+a_{0}V+\lambda _{1}\mathcal{I}%
_{1}+\lambda _{2}\mathcal{I}_{2},1-V_{E}\right\} =0,  \label{Eq121}
\end{equation}%
where%
\begin{equation}
\mathcal{I}_{i}\left( t,E\right) =\delta _{i}\int_{0}^{E}V\left(
t,E-J_{i}\right) e^{-\delta _{i}J_{i}}dJ_{i}=\delta _{i}\int_{0}^{E}V\left(
t,j\right) e^{-\delta _{i}\left( E-j\right) }dj,\ \ \ i=1,2.  \label{Eq122}
\end{equation}%
}Symbolically, we can represent Eq. (\ref{Eq121}) in the form 
\begin{equation}
\max \left\{ V_{t}+\mathcal{L}\left( V\right) ,1-V_{E}\right\} =0,
\label{Eq123}
\end{equation}%
where%
\begin{equation}
\mathcal{L}\left( V\right) =a_{2}V_{EE}+a_{1}V_{E}+a_{0}V+\lambda _{1}%
\mathcal{I}_{1}+\lambda _{2}\mathcal{I}_{2}.  \label{Eq124}
\end{equation}

Solution $V\left( t,E\right) $ of this variational inequality cannot be
computed analytically and has to be determined numerically. To this end, we
use the method proposed by Lipton (2003) and replace the variational
inequality in question by the following one%
\begin{equation}
\begin{array}{c}
\max \left\{ -V_{\tau }+a_{2}V_{EE}+a_{1}V_{E}+a_{0}V+\lambda _{1}\mathcal{I}%
_{1}+\lambda _{2}\mathcal{I}_{2},1-V_{E}\right\} =0, \\ 
\mathcal{I}_{i,E}+\delta _{i}\mathcal{I}_{i}-\delta _{i}V=0, \\ 
V\left( 0,E\right) =E, \\ 
V\left( \tau ,0\right) =0,%
\end{array}
\label{Eq125}
\end{equation}%
where $\tau =T-t$. The corresponding problem is solved in a relatively
straightforward way by computing $\mathcal{I}_{i}$ and performing the
operation $\max \left\{ .,.\right\} $ explicitly, while calculating $V$ in
the usual Crank-Nicolson manner. The corresponding solution is shown in
Figure \ref{Fig12}.

\begin{equation*}
\text{Figure \ref{Fig12} near here.}
\end{equation*}%
{\ }

For the $T\rightarrow \infty $ limit, the time-independent maximization
problem has the form{%
\begin{equation}
\begin{array}{c}
\max \left\{ \mathcal{L}\left( V\right) ,1-V_{E}\right\} =0, \\ 
V\left( 0\right) =0,%
\end{array}
\label{Eq126}
\end{equation}%
}or, equivalently,{%
\begin{equation}
\begin{array}{c}
\mathcal{L}\left( V\right) \left( E\right) =0,\ \ \ 0<E\leq E^{\ast }, \\ 
V\left( E\right) =E+V\left( E^{\ast }\right) -E^{\ast },\ \ \ E^{\ast
}<E<\infty , \\ 
V\left( 0\right) =0, \\ 
V_{E}\left( E^{\ast }\right) =1, \\ 
\ V_{EE}\left( E^{\ast }\right) =0.%
\end{array}%
\   \label{Eq127}
\end{equation}%
Here $E^{\ast }$ is not known in advance and has to be determined as part of
the calculation.}

It turns out that the time-independent problem can be solved analytically. {%
Since we are dealing with a Levy process, }we\ {have%
\begin{equation}
\mathcal{L}\left( e^{\xi E}\right) =\Psi \left( \xi \right) e^{\xi E}-\frac{%
\lambda _{1}\delta _{1}}{\xi +\delta _{1}}e^{-\delta _{1}E}-\frac{\lambda
_{2}\delta _{2}}{\xi +\delta _{2}}e^{-\delta _{2}E},  \label{Eq128}
\end{equation}%
where }$\Psi \left( \xi \right) $ is the symbol of the pseudo-differential
operator $\mathcal{L}$,{%
\begin{equation}
\Psi \left( \xi \right) =a_{2}\xi ^{2}+a_{1}\xi +a_{0}+\frac{\lambda
_{1}\delta _{1}}{\xi +\delta _{1}}+\frac{\lambda _{2}\delta _{2}}{\xi
+\delta _{2}}.  \label{Eq129}
\end{equation}%
Denote by $\xi _{j}$, $j=1,...,4$, the roots of the (polynomial) equation%
\begin{equation}
\Psi \left( \xi \right) =0.  \label{Eq130}
\end{equation}%
}The corresponding function $\Psi \left( \xi \right) $ for a representative
set of parameters is exhibited in Figure \ref{Fig13}, which clearly shows
that all roots of Eq. (\ref{Eq130}) are real.

\begin{equation*}
\text{Figure \ref{Fig13} near here.}
\end{equation*}%
{\ Then a linear combination%
\begin{equation}
V\left( E\right) =\sum_{j}C_{j}e^{\xi _{j}E},  \label{Eq131}
\end{equation}%
solves the pricing problem and the boundary conditions (\ref{Eq127})
provided that}%
\begin{equation}
\left( 
\begin{array}{cccc}
1 & 1 & 1 & 1 \\ 
\left( \xi _{1}+\delta _{1}\right) ^{-1} & \left( \xi _{2}+\delta
_{1}\right) ^{-1} & \left( \xi _{3}+\delta _{1}\right) ^{-1} & \left( \xi
_{4}+\delta _{1}\right) ^{-1} \\ 
\left( \xi _{1}+\delta _{2}\right) ^{-1} & \left( \xi _{2}+\delta
_{2}\right) ^{-1} & \left( \xi _{3}+\delta _{2}\right) ^{-1} & \left( \xi
_{4}+\delta _{2}\right) ^{-1} \\ 
\xi _{1}e^{\xi _{1}E^{\ast }} & \xi _{2}e^{\xi _{2}E^{\ast }} & \xi
_{3}e^{\xi _{3}E^{\ast }} & \xi _{4}e^{\xi _{4}E^{\ast }} \\ 
\xi _{1}^{2}e^{\xi _{1}E^{\ast }} & \xi _{2}^{2}e^{\xi _{2}E^{\ast }} & \xi
_{3}^{2}e^{\xi _{3}E^{\ast }} & \xi _{4}^{2}e^{\xi _{4}E^{\ast }}%
\end{array}%
\right) \left( 
\begin{array}{c}
C_{1} \\ 
C_{2} \\ 
C_{3} \\ 
C_{4}%
\end{array}%
\right) =\left( 
\begin{array}{c}
0 \\ 
0 \\ 
0 \\ 
1 \\ 
0%
\end{array}%
\right) .  \label{Eq132}
\end{equation}%
Eqs (\ref{Eq132}) should be thought of as a system of five equations for
five unknowns, namely, $\left( C_{1},C_{2},C_{3},C_{4}\right) $ and $E^{\ast
}$. The corresponding profile $V\left( E\right) $ is presented in Figure \ref%
{Fig14}.

\begin{equation*}
\text{Figure \ref{Fig14} near here.}
\end{equation*}%
{\ }

This graph shows that on the interval $\left[ 0,E^{\ast }\right) $ we have $%
V_{E}>1$. Accordingly, the coefficient $\left( 1-V_{E}\right) $ in front of $%
d$ in Eq. (\ref{Eq117}) is negative, so that the optimal $d$ has to be zero.
To put it differently, it is optimal for the bank not to pay any dividends
until $E$\ reaches the optimal level $E^{\ast }$. On the interval $\left(
E^{\ast },\infty \right) $ we have $V_{E}>1$, so that $d$ is not determined.
However, this is not particularly important, since when $E$\ exceeds the
optimal level $E^{\ast }$ it is optimal to pay all the excess equity in
dividends. This situation occurs because we allow for infinite dividend
rate, and hence lump-sum payments. When $d$ is bounded, the corresponding
optimization problem is somewhat different, but can still be solved along
similar lines.

Comparison of Figures \ref{Fig14}(a) and \ref{Fig14}(b) shows that $V\left(
E\right) $ is an excellent approximation for $V\left( T,E\right) $ for
longer maturities $T$.

\section{Conclusions\label{Conclusions}}

In this paper we proposed a simple and consistent theory that enables one to
examine the banking system at three levels of granularity, namely, as a
whole, as an interconnected collection of banks with mutual liabilities;
and, finally, as an individual bank. We demonstrated that the banking system
plays a pivotal role in the monetary circuit context and is necessary for
the success of the economy. Even in a relatively simple context we gained
some nontrivial insights into money creation by banks and its consequences,
including naturally occurring interbank linkages, as well as the role of
multiple constraints banks are operating under.

The consistent quantitative description of the monetary circuit in
continuous time became possible after the introduction of stochastic
consumption by rentiers into the model, which enabled us to reconcile the
equations with economic reality. We built a quantitative description of the
monetary circuit\ that can be calibrated to real macro economic data which
we solved mathematically. The developed framework can be further expanded by
adding various sectors of the economy. It is clear that more\ advanced
models will naturally provide deeper actionable insights, which can be used
for a variety of purposes, such as setting the monetary policy, positioning
banks for responsible growth, and macro investing.

At the top level, we considered the banking system as a whole, disguising
therefore the structure of the banking sector and precluding investigation
of defaults within it. It is hard to overestimate the importance of the
quantitative approach that enables the description of a possible chain of
events in the interconnected banking system in the aftermath of the crisis
of 2007-2009! Hence, we expanded our analysis to the intermediate level, and
demonstrated how the asset-liability balancing act creates nontrivial
linkages between various banks. We used techniques developed for credit
default pricing to show that these\ linkages can cause unexpected
instabilities in the overall system. Our model can be expanded in several
directions, for instance, by incorporating interbank derivatives, such as
swaps, into the picture. It can provide insights into snowball effects
associated with multiple simultaneous (or almost simultaneous) defaults in
the banking system.

Finally, viewed at the bottom level, banks, as all other corporations, have
a fiduciary obligation to responsibly maximize their profitability. Given
the specifics of the banking business, such a maximization of profitability
is intrinsically linked to balance sheet optimization, which is used in
order to choose an optimal mix of assets and liabilities. We formulated the\
constrained optimization problem in the most general case, as well as its
reduced version in a specific case of the equity part of the capital
structure. Although simplified, the reduced problem still includes such
salient elements of the equity dynamics as liquidity and solvency jumps. We
then proposed a scheme to efficiently solve the corresponding constrained
optimization problem.

We hope that our theory of MMC will stimulate further research along the
lines suggested in the paper. In particular, to help to predict future
economic crises, which naturally arise within the proposed framework.

\section*{Acknowledgments}
The author is grateful to Russell Barker, Agostino Capponi, Michael
Dempster, Andrew Dickinson, Darrel Duffie, Paul Glasserman, Tom Hurd, Andrey
Itkin, Marsha Lipton, and Rajeev Virmani for useful conversations. This
paper was presented at Bloomberg Quant Seminar Series in NY, Global
Derivatives Conference in Amsterdam, Workshop on Models and Numerics in
Financial Mathematics at the Lorentz Center in Leiden, Workshop on Systemic
Risk at Columbia University in NY, and 7th General Advanced Mathematical
Methods in Finance and Swissquote Conference in Lausanne. Feedback and
suggestions from participants in these events are much appreciated. The
invaluable help of Marsha Lipton in bringing this work to fruition and
preparing it for publication cannot be overestimated.

\section{Appendix A\label{AppendixA}}

To make our calculations in Section \ref{BankingSystem} more concrete, let
us consider the case of just two banks with mutual obligations without
netting, $N=2$. Additional details can be found in Itkin and Lipton (2015b).

For $0<t<T$ default boundaries have the form%
\begin{equation}
A_{i}\leq \Lambda _{i}=\left\{ 
\begin{array}{cc}
R_{i}\left( L_{i}+L_{i\bar{\imath}}\right) -L_{\bar{\imath}i}\equiv \Lambda
_{i}^{<},\ \  & \ t<T, \\ 
L_{i}+L_{i\bar{\imath}}-L_{\bar{\imath}i}\equiv \Lambda _{i}^{=},\ \  & t=T,%
\end{array}%
\right.  \label{EqA1}
\end{equation}%
\begin{equation}
A_{i}\leq \tilde{\Lambda}_{i}=\left\{ 
\begin{array}{cc}
R_{i}\left( L_{i}+L_{i\bar{\imath}}-R_{\bar{\imath}}L_{\bar{\imath}i}\right)
\equiv \tilde{\Lambda}_{i}^{<}, & t<T, \\ 
L_{i}+L_{i\bar{\imath}}-R_{\bar{\imath}}L_{\bar{\imath}i}\equiv \tilde{%
\Lambda}_{i}^{=},\ \ \  & t=T.%
\end{array}%
\right.  \label{EqA2}
\end{equation}%
where $\bar{\imath}=3-i$. In the $\left( A_{1},A_{2}\right) $ quadrant we
have four domains%
\begin{eqnarray}
\mathcal{D}\left( 1,1\right) &=&\left\{ A_{1}>\Lambda _{1}^{=},\ \ \
A_{2}>\Lambda _{2}^{=}\right\} ,  \label{EqA3} \\
\mathcal{D}\left( \delta _{i,1},\delta _{i,2}\right) &=&\left\{ A_{i}>\frac{%
\Delta -L_{\bar{\imath}i}A_{\bar{\imath}}}{L_{\bar{\imath}}+L_{\bar{\imath}i}%
},\ \ \ \Lambda _{\bar{\imath}}^{<}<A_{\bar{\imath}}<\Lambda _{\bar{\imath}%
}^{=}\right\} ,\ \ \ i=1,2,  \notag \\
\mathcal{D}\left( 0,0\right) &=&\left\{ A_{1}>\Lambda _{1}^{<},A_{2}>\Lambda
_{2}^{<}\right\} -\mathcal{D}\left( 1,1\right) -\mathcal{D}\left( 1,0\right)
-\mathcal{D}\left( 0,1\right) ,  \notag
\end{eqnarray}%
where $\delta _{i,j}$ is the Kronecker delta, and 
\begin{equation}
\Delta =L_{1}L_{2}+L_{1}L_{21}+L_{2}L_{12}.  \label{EqA4}
\end{equation}%
It is clear that in $\mathcal{D}\left( 1,1\right) $ both banks survive, in $%
\mathcal{D}\left( 1,0\right) $ the first bank survives and the second
defaults, in $\mathcal{D}\left( 0,1\right) $ the second bank survives and
the first defaults, and in $\mathcal{D}\left( 0,0\right) $ both banks
default. The corresponding domains are shown in Figure \ref{Fig15}(a).

In log coordinates the domain $\mathcal{D}_{i}$ has the form%
\begin{equation}
\mathcal{D}\left( \delta _{i,1},\delta _{i,2}\right) =\left\{ X_{i}>\Theta
_{i}\left( X_{\bar{\imath}}\right) ,0<X_{\bar{\imath}}<M_{\bar{\imath}%
}^{=}\right\} ,  \label{EqA5}
\end{equation}%
where%
\begin{equation}
\Theta _{i}\left( X_{\bar{\imath}}\right) =\sqrt{\frac{\sigma _{\bar{\imath}}%
}{\sigma _{i}}}\ln \left( \frac{\Delta -L_{\bar{\imath}i}\left( R_{\bar{%
\imath}}\left( L_{\bar{\imath}}+L_{\bar{\imath}i}\right) -L_{i\bar{\imath}%
}\right) \exp \left( \sqrt{\sigma _{\bar{\imath}}/\sigma _{i}}X_{\bar{\imath}%
}\right) }{\left( R_{i}\left( L_{i}+L_{i\bar{\imath}}\right) -L_{\bar{\imath}%
i}\right) \left( L_{\bar{\imath}}+L_{\bar{\imath}i}\right) }\right) .
\label{EqA6}
\end{equation}%
We emphasize that the domain $\mathfrak{D}_{i}$ has a curvilinear boundary
which depends on the value of $A_{i}$. It is worth noting that 
\begin{equation}
\Theta _{i}\left( 0\right) =\tilde{M}_{i}^{=},\ \ \ \ \ \Theta _{i}\left(
\mu _{\bar{\imath}}^{=}\right) =M_{i}^{=}.  \label{EqA7}
\end{equation}%
The corresponding domains are shown in Figure \ref{Fig15}(b).

\begin{equation*}
\text{Figure \ref{Fig15} near here.}
\end{equation*}%
{\ }

Payoffs for different options are as follows.\ For the joint survival
probability%
\begin{eqnarray}
Q\left( T,A_{1},A_{2}\right) &=&\boldsymbol{1}_{\left( A_{1},A_{2}\right)
\in \mathcal{D}\left( 1,1\right) },  \label{EqA8} \\
Q\left( t,\delta _{i,1}\Lambda _{i}^{<}+\delta _{i,2}A_{\bar{\imath}},\delta
_{i,2}\Lambda _{i}^{<}+\delta _{i,1}A_{\bar{\imath}}\right) &=&0,\ \ \ i=1,2.
\notag
\end{eqnarray}%
For marginal survival probabilities%
\begin{equation}
Q_{i}\left( T,A\,_{1},A_{2}\right) =\boldsymbol{1}_{\left(
A_{1},A_{2}\right) \in \mathcal{D}\left( 1,1\right) +\mathcal{D}\left(
\delta _{i,1},\delta _{i,2}\right) },  \label{EqA9}
\end{equation}%
For the CDSs on the first and second bank the payoffs are as follows%
\begin{equation}
C_{i}\left( T,A\,_{1},A_{2}\right) =\left\{ 
\begin{array}{cc}
0, & \left( A_{1},A_{2}\right) \in \mathcal{D}\left( 1,1\right) +\mathcal{D}%
\left( \delta _{i,1},\delta _{i,2}\right) , \\ 
1-\frac{A_{i}+L_{\bar{\imath}i}}{L_{i}+L_{i\bar{\imath}}}, & \left(
A_{1},A_{2}\right) \in \mathcal{D}\left( \delta _{\bar{\imath},1},\delta _{%
\bar{\imath},2}\right) , \\ 
1-\frac{A_{i}+\varkappa _{\bar{\imath}}L_{\bar{\imath}i}}{L_{i}+L_{i\bar{%
\imath}}}, & \left( A_{1},A_{2}\right) \in \mathcal{D}\left( 0,0\right) ,%
\end{array}%
\right.  \label{EqA10}
\end{equation}%
where the coefficients $\varkappa _{i}$ are determined from the detailed
balance equations%
\begin{eqnarray}
A_{1}+\varkappa _{2}L_{21} &=&\varkappa _{1}\left( L_{1}+L_{12}\right) ,
\label{EqA11} \\
A_{2}+\varkappa _{1}L_{12} &=&\varkappa _{2}\left( L_{2}+L_{21}\right) , 
\notag
\end{eqnarray}%
so that%
\begin{equation}
\varkappa _{i}=\frac{L_{\bar{\imath}}A_{i}+L_{\bar{\imath}i}\left( A_{i}+A_{%
\bar{\imath}}\right) }{\Delta }.  \label{EqA12}
\end{equation}%
Finally, for the FTD the payoff has the form%
\begin{equation}
F\left( T,A\,_{1},A_{2}\right) =\left\{ 
\begin{array}{cc}
0, & \left( A_{1},A_{2}\right) \in \mathcal{D}\left( 1,1\right) , \\ 
1-\frac{A_{\bar{\imath}}+L_{i\bar{\imath}}}{L_{\bar{\imath}}+L_{\bar{\imath}%
i}} & \left( A_{1},A_{2}\right) \in \mathcal{D}\left( \delta _{i,1},\delta
_{i,2}\right) , \\ 
\max \left\{ 1-\frac{A_{1}+\varkappa _{2}L_{21}}{L_{1}+L_{12}},1-\frac{%
A_{2}+\varkappa _{1}L_{12}}{L_{2}+L_{21}}\right\} , & \left(
A_{1},A_{2}\right) \in \mathcal{D}\left( 0,0\right) ,%
\end{array}%
\right.  \label{EqA13}
\end{equation}

For brevity, we consider just the calculation of the joint and marginal
survival probabilities. The joint survival probability $Q\left(
t,X_{1},X_{2}\right) $ solves the following terminal boundary value problem%
\begin{gather}
Q_{t}\left( t,X_{1},X_{2}\right) +\mathcal{L}Q\left( t,X_{1},X_{2}\right) =0,
\label{EqA14} \\
Q\left( T,X_{1},X_{2}\right) =\boldsymbol{1}_{X\in \mathcal{D}\left(
1,1\right) },  \notag \\
Q\left( t,X_{1},0\right) =0,\ \ \ Q\left( t,0,X_{2}\right) =0,  \notag
\end{gather}%
The corresponding marginal survival probability for the first bank, say, $%
Q_{1}\left( t,X_{1},X_{2}\right) $, which is a function of \emph{both} $%
X_{1} $ and $X_{2}$ solves the following terminal boundary value problem%
\begin{gather}
Q_{1,t}\left( t,X_{1},X_{2}\right) +\mathcal{L}Q_{1}\left(
t,X_{1},X_{2}\right) =0,  \label{EqA15} \\
Q_{1}\left( T,X_{1},X_{2}\right) =\boldsymbol{1}_{X\in \mathfrak{D}\left(
1,1\right) }+\boldsymbol{1}_{X\in \mathfrak{D}\left( 1,0\right) },  \notag \\
Q_{1}\left( t,0,X_{2}\right) =0,  \notag \\
Q_{1}\left( t,X_{1},0\right) =\left\{ 
\begin{array}{cc}
q_{1}\left( t,X_{1}\right) , & X_{1}\geq M_{1}^{\left( 2\right) ,<}, \\ 
0, & X_{1}<M_{1}^{\left( 2\right) ,<},%
\end{array}%
\right. ,\ \   \notag
\end{gather}

Here $q_{1}\left( t,X_{1}\right) $ is the 1D survival probability, which
solves the following terminal boundary value problem%
\begin{gather}
q_{1,t}\left( t,X_{1}\right) +\frac{1}{2}q_{1,X_{1}X_{1}}+\xi
_{1}q_{1,X_{1}}=0,  \label{EqA16} \\
q_{1}\left( T,X_{1}\right) =\boldsymbol{1}_{\left\{ X_{1}>M_{1}^{\left(
2\right) ,=}\right\} },  \notag \\
q_{1}\left( t,M_{1}^{\left( 2\right) ,<}\right) =0,\ \ \   \notag
\end{gather}%
It is very easy to show that%
\begin{eqnarray}
q_{1}\left( t,X_{1}^{\prime }\right) &=&N\left( -\frac{M_{1}^{\left(
2\right) ,=}-X_{1}^{\prime }-\xi _{1}\tau }{\sqrt{\tau }}\right)
\label{EqA17} \\
&&-e^{-2\xi _{1}\left( X_{1}^{\prime }-M_{1}^{\left( 2\right) ,<}\right)
}N\left( -\frac{M_{1}^{\left( 2\right) ,=}+X_{1}^{\prime }-2M_{1}^{\left(
2\right) ,<}-\xi _{1}\tau }{\sqrt{\tau }}\right) ,  \notag
\end{eqnarray}%
where $\tau =T-t$.

The corresponding 2D Green's function has the form (see, e.g., Lipton 2001,
Lipton and Savecu 2014):%
\begin{equation}
\begin{array}{c}
G\left( t,X_{1},X_{2}\right) =e^{-\left( \theta ,\xi \right) t/2+\theta
\cdot \left( X-X^{\prime }\right) }\bar{G}\left( t,X_{1},X_{2}\right) , \\ 
\bar{G}\left( t,X_{1},X_{2}\right) =\frac{2e^{-\left( R^{2}+R^{\prime
2}\right) /2t}}{\bar{\rho}\varpi t}\dsum\limits_{n=1}^{\infty }I_{\nu
_{n}}\left( \frac{RR^{\prime }}{t}\right) \sin \left( \nu _{n}\phi \right)
\sin \left( \nu _{n}\phi ^{\prime }\right) ,%
\end{array}
\label{EqA18}
\end{equation}%
where%
\begin{equation}
\begin{array}{c}
C=\left( 
\begin{array}{cc}
1 & \rho \\ 
\rho & 1%
\end{array}%
\right) ,\ \ \ \ \ C^{-1}=\frac{1}{\bar{\rho}^{2}}\left( 
\begin{array}{cc}
1 & -\rho \\ 
-\rho & 1%
\end{array}%
\right) , \\ 
\theta =C^{-1}\xi ,\ \ \ \bar{\rho}=\sqrt{1-\rho ^{2}}, \\ 
\varpi =\arctan \left( -\frac{\bar{\rho}}{\rho }\right) ,\ \ \ \ \ \nu _{n}=%
\frac{n\pi }{\varpi }>n, \\ 
R=\sqrt{\left( C^{-1}X,X\right) },\ \ \ \ \ R^{\prime }=\sqrt{\left(
C^{-1}X^{\prime },X^{\prime }\right) }, \\ 
\phi =\arctan \left( \frac{\bar{\rho}X_{1}}{-\rho X_{1}+X_{2}}\right) ,\ \ \
\phi ^{\prime }=\arctan \left( \frac{\bar{\rho}X_{1}^{\prime }}{-\rho
X_{1}^{\prime }+X_{2}^{\prime }}\right) .%
\end{array}
\label{EqA19}
\end{equation}%
It is clear that%
\begin{equation}
\begin{array}{c}
G_{X_{2}}\left( t,X_{1},0\right) =e^{-\left( \theta ,\xi \right) t/2+\theta
_{1}X_{1}-\theta \cdot X^{\prime }}\bar{G}_{X_{2}}\left( t,X_{1},0\right) ,
\\ 
\bar{G}_{X_{2}}\left( t,X_{1},0\right) =\frac{2e^{-\left( X_{1}^{2}/\bar{\rho%
}^{2}+R^{\prime 2}\right) /2t}}{\varpi tX_{1}}\dsum\limits_{n=1}^{\infty
}\left( -1\right) ^{n+1}\nu _{n}I_{\nu _{n}}\left( \frac{X_{1}R^{\prime }}{%
\bar{\rho}t}\right) \sin \left( \nu _{n}\phi ^{\prime }\right) .%
\end{array}
\label{EqA20}
\end{equation}%
Substitution of these formulas in Eqs (\ref{Eq89}), (\ref{Eq90a}) yield
semi-analytical expressions for $Q$ and $Q_{1}$. The corresponding
expression for $Q_{2}$ is similar.

We present $Q_{1}\left( 0,X_{1},X_{2}\right) $ and the difference $%
q_{1}\left( 0,X_{1}\right) -Q_{1}\left( 0,X_{1},X_{2}\right) $ in Figures {%
\ref{Fig16}(a) and \ref{Fig16}(b), respectively. These Figures show that
mutual obligations significantly impact survival probabilities and other
quantities of interest.}

\begin{equation*}
\text{Figure \ref{Fig16} near here.}
\end{equation*}


{\ 
\begin{figure}[t]
{\ \center%
\includegraphics[width=1.0\textwidth, angle=0]{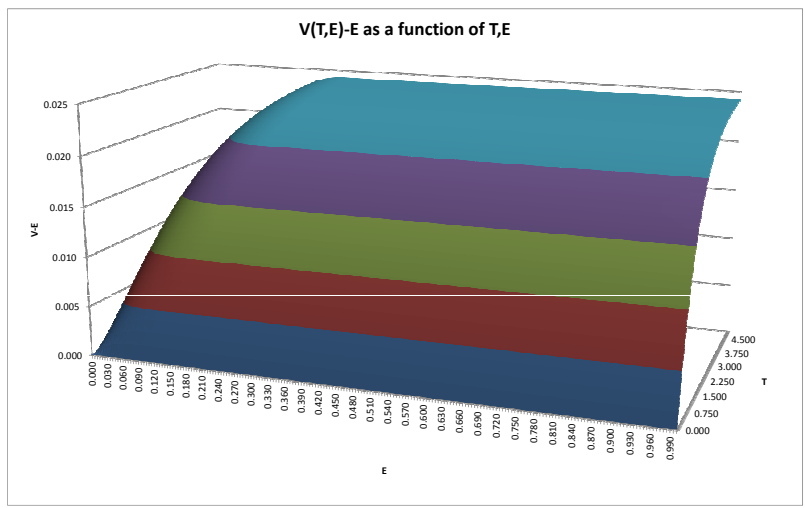}}
\caption{Excess value $V\left( T,E\right) -E$ viewed as a function of time
to maturity $T$ and equity value $E$ for some representative parameters: $%
\protect\sigma =0.25$, $\protect\mu =0.05$, $\protect\nu =0.10$, $\protect%
\lambda _{1}=0.05$, $\protect\delta _{1}=3.00$, $\protect\lambda _{2}=0.02$, 
$\protect\delta _{2}=1.00$. }
\label{Fig12}
\end{figure}
\ 
\begin{figure}[t]
{\ \center%
\includegraphics[width=1.0\textwidth, angle=0]{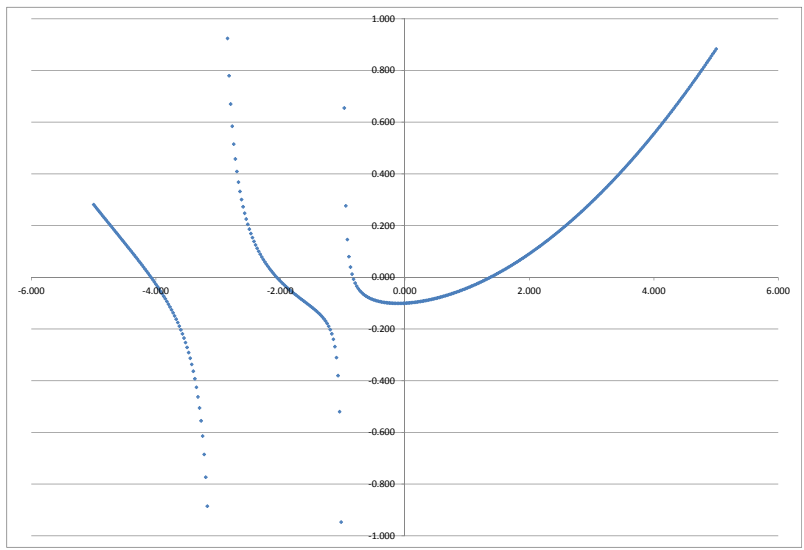}} 
\caption{Function $\Psi \left( \protect\xi \right) $ for the same set of
parameters as used in Figure \protect\ref{Fig12}. It is easy to see that
equation $\Psi \left( \protect\xi \right) =0$ has four roots $\protect\xi %
_{1}=-4.08$, $\protect\xi _{2}=-2.06$, $\protect\xi _{3}=-0.84$, and $%
\protect\xi _{4}=1.37$.}
\label{Fig13}
\end{figure}
}

{\ }%
\begin{sidewaysfigure}[t]
\centering%
\makebox[\textwidth]{
  \subfigure[]
  {
  	\includegraphics[width = 0.5\textwidth]{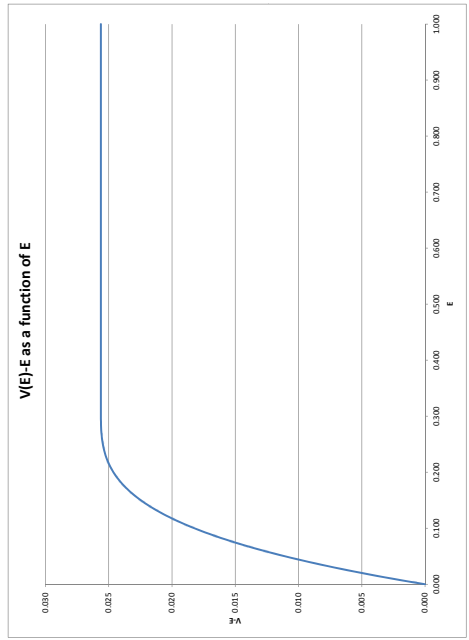}
  }
  \subfigure[]
  {
  	\includegraphics[width = 0.5\textwidth]{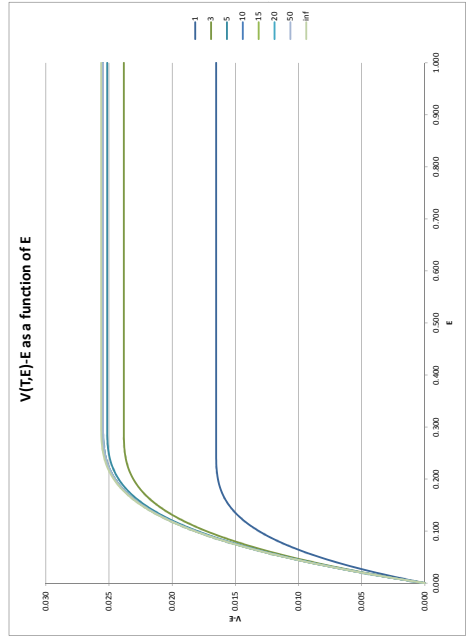}
  }
}
\caption{Excess value $V(T,E)-E$ viewed as a function of equity value $E$ for different times $T$(subfigure a); 
limiting excess value $V(\inf,E)-E$ viewed as a function of equity value $E$ (subfigure b). 
Subfigures (a) and (b) agree perfectly. The same parameters as in Figure \ref{Fig12} are used.}
\label{Fig14}
\end{sidewaysfigure}{\ }

{\ }\clearpage%
\begin{sidewaysfigure}[t]
\centering%
\makebox[\textwidth]{
  \subfigure[]
  {
  	\includegraphics[width = 0.5\textwidth]{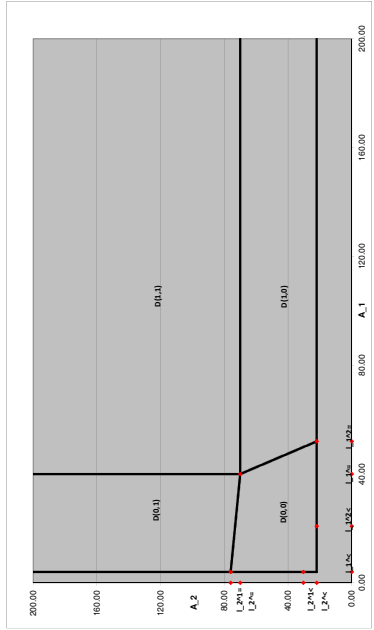}
  }
  \subfigure[]
  {
  	\includegraphics[width = 0.5\textwidth]{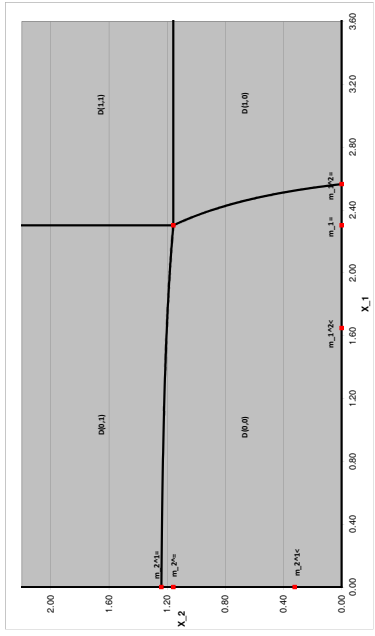}
  }
}
\caption{Default boundaries for two interconnected banks in the $(A_1,A_2)$ plane (subfigure a), 
and in the $(X_1,X_2)$ plane (subfigure b). Here $L_1=50, L_12=10, R_1=0.4, L_2=60, L_21=20, R_2=0.4$.}
\label{Fig15}
\end{sidewaysfigure}

\clearpage%
\begin{sidewaysfigure}[t]
\centering%
\makebox[\textwidth]{
  \subfigure[]
  {
  	\includegraphics[width = 0.5\textwidth]{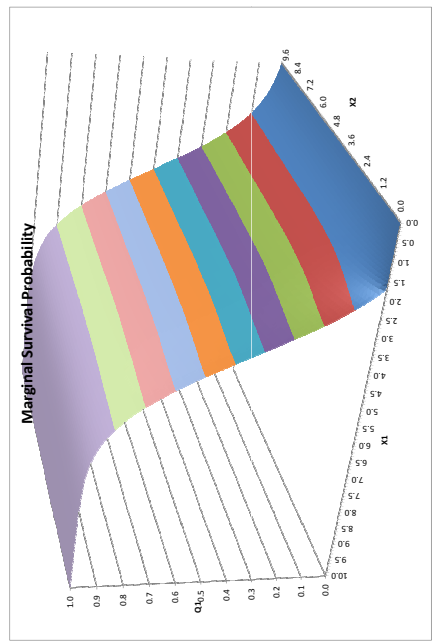}
  }
  \subfigure[]
  {
  	\includegraphics[width = 0.5\textwidth]{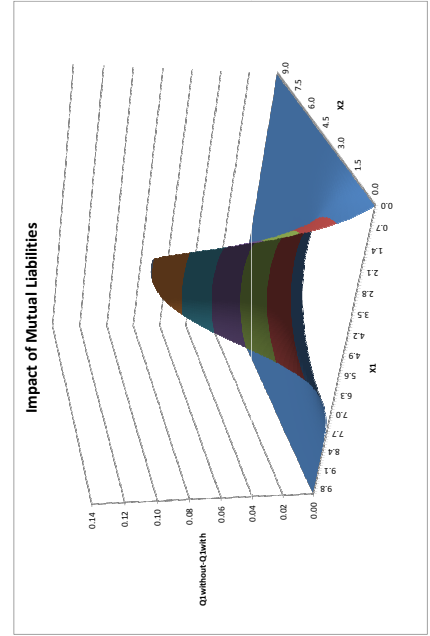}
  }
}
\caption{Marginal survival probability $Q_1(0,X_1,X_2)$ (subfigure a); 
decrease in marginal survival probability $q_1(0,X_1)-Q_1(0,X_1,X_2)$ due to mutual liabilities between two banks (subfigure b).
We use the same parameters as in Figure \ref{Fig15}. 
In addition, we choose $\sigma_1=\sigma_2=0.4, \rho_{12}=0, \mu=0, T=12.5 Y$.}
\label{Fig16}
\end{sidewaysfigure}

\end{document}